\documentclass[twocolumn]{aastex7}

\let\tablewidth\relax
\usepackage{tabularray}
\usepackage{verbatim}
\usepackage{tabularx}
\usepackage{array}
\usepackage{comment}
\usepackage{ulem}
\usepackage{xcolor}
\usepackage{graphicx}   
\usepackage{array}
\usepackage{multirow}
\usepackage{amsmath}
\usepackage{lineno}
\usepackage{subcaption}
\usepackage{float}
\usepackage{afterpage}
\usepackage{booktabs}
\usepackage{siunitx}
\usepackage{multirow} 

\graphicspath{{./}{figures/}}

\shorttitle{impact of resolution in simulations of galaxy}
\shortauthors{Li, Zhu, Wang, Chen, \& Feng.}

\begin{document}

\title{Toward Resolution Convergence for Star Formation and Galactic Winds in Galaxy Simulations}

\author{Xue-Fu Li}
\email{lixf96@mail2.sysu.edu.cn}
\affil{School of Physics and Astronomy, Sun Yat-Sen University, Zhuhai campus, No. 2, Daxue Road \\
Zhuhai, Guangdong, 519082, China}
\affil{CSST Science Center for the Guangdong-Hong Kong-Macau Greater Bay Area, Daxue Road 2, 519082, Zhuhai, China}

\author[0000-0002-1189-2855]{Weishan Zhu}
\email{zhuwshan5@mail.sysu.edu.cn}
\affil{School of Physics and Astronomy, Sun Yat-Sen University, Zhuhai campus, No. 2, Daxue Road \\
Zhuhai, Guangdong, 519082, China}
\affil{CSST Science Center for the Guangdong-Hong Kong-Macau Greater Bay Area, Daxue Road 2, 519082, Zhuhai, China}

\author{Tian-Rui Wang}
\email{wangtr3@alumni.sysu.edu.cn}
\affil{School of Physics and Astronomy, Sun Yat-Sen University, Zhuhai campus, No. 2, Daxue Road \\
Zhuhai, Guangdong, 519082, China}
\affil{CSST Science Center for the Guangdong-Hong Kong-Macau Greater Bay Area, Daxue Road 2, 519082, Zhuhai, China}

\author{Huan Chen}
\email{chenh699@mail2.sysu.edu.cn}
\affil{School of Physics and Astronomy, Sun Yat-Sen University, Zhuhai campus, No. 2, Daxue Road \\
Zhuhai, Guangdong, 519082, China}
\affil{CSST Science Center for the Guangdong-Hong Kong-Macau Greater Bay Area, Daxue Road 2, 519082, Zhuhai, China}

\author{Long-Long Feng}
\email{flonglong@mail.sysu.edu.cn}
\affil{School of Physics and Astronomy, Sun Yat-Sen University, Zhuhai campus, No. 2, Daxue Road \\
Zhuhai, Guangdong, 519082, China}
\affil{CSST Science Center for the Guangdong-Hong Kong-Macau Greater Bay Area, Daxue Road 2, 519082, Zhuhai, China}

\correspondingauthor{Weishan Zhu}
\email{zhuwshan5@mail.sysu.edu.cn}



\begin{abstract}
The predictive power of hydrodynamic simulations of galaxies is often hampered by the resolution dependence of sub-grid models for star formation (SF) and stellar feedback. We present three-dimensional hydrodynamic simulations of an M82-like low-mass starburst galaxy at resolutions of $4$--$32$ pc and develop optimized sub-grid prescriptions to mitigate these numerical inconsistencies. For SF, we adopt a $\sim20$ pc physical-scale control volume, corresponding to the typical size of giant molecular clouds (GMCs), for gas accretion onto sink particles, reducing its dependence on numerical resolution. We further incorporate an empirical, resolution- and density-dependent SF scheme that accounts for variations in gas surface density across spatial scales, as suggested by observations. Furthermore, we refine the supernova feedback coupling by anchoring the injection volume to the physical cooling radius of the supernova remnant ($R_{\rm cool}$) and utilizing a resolution-dependent thermal-to-kinetic energy partition. These improvements substantially enhance convergence across $4$--$32$ pc: variations in the total stellar mass formed and kinetic energy are reduced to $\sim20\%$, while relative errors in the cool-phase mass outflow rate and integrated X-ray luminosity decrease to below $\sim40\%$. The warm and hot outflow phases remain less well converged, with variations of $\sim50\%$ and $\sim75\%$, respectively. The upgraded model also improves the convergence of normalized wind properties, with relative errors below $\sim50\%$, $\sim80\%$, and $\sim20\%$ for the time-averaged mass loading factor, energy loading factor, and X-ray emission efficiency, respectively. Our results provide a practical framework for achieving more robust simulations of starburst galaxies.
\end{abstract}

\keywords{\uat{Galaxy evolution}{594}; \uat{Starburst galaxies}{1570}; \uat{Stellar feedback}{1602};\uat{Galactic winds}{572}; \uat{Hydrodynamical
simulations}{767}}



\section{Introduction} \label{sec:intro}
Beyond the background gravitational potential, which is primarily dominated by dark matter, baryonic processes, including radiative gas cooling, star formation (SF), and feedback from both stellar populations and active galactic nuclei (AGN), are essential components of modern galaxy formation and evolution theory. Despite their importance, several key aspects of baryonic physics remain poorly understood or require deeper investigation. These include the mechanisms of gas accretion, the regulation of star formation activity, and the complex baryon cycling occurring across galaxy-halo scales.

A critical element of this baryonic cycle is the multiphase, galaxy-scale outflow powered by stellar or AGN feedback, commonly referred to as a galactic wind. Over the past six decades, galactic winds have been extensively observed in the local universe \citep{1963ApJ...137.1005L, 1990ApJS...74..833H, 1999ApJ...513..156M, 2005ApJS..160..115R, 2015ApJ...809..147H, 2023arXiv231006614X}, as well as in galaxies at intermediate and high redshifts \citep[e.g.,][]{2009ApJ...692..187W, 2012ApJ...760..127M, 2023ApJ...959..124K, 2023ApJ...949....9P, 2024A&A...685A..99C, 2024MNRAS.531.4560W}. In this work, we focus specifically on galactic outflows driven by stellar feedback.

Galactic outflows regulate star formation efficiency and can even lead to quenching, providing a natural mechanism to reconcile the observed galaxy luminosity function with the predicted dark matter halo mass function (e.g., the "missing satellites" problem; \citealt{1986ApJ...303...39D, 1994MNRAS.271..781C, 1999ApJ...519L.109C, 2003ApJ...599...38B, 2005ARA&A..43..769V, 2017ARA&A..55...59N}). Simultaneously, these winds transport heavy elements into the circumgalactic (CGM) and intergalactic mediums (IGM).

Despite their important role in galaxy evolution, many questions regarding the properties and physical mechanisms of these outflows remain open. Theoretically, the interplay between star formation, feedback and interstellar medium (ISM) is highly nonlinear, and the processes determining the properties of the multiphase gas within the outflow are still debated.  Observationally, significant uncertainties persist in the inferred density profiles, velocities, and mass-loading rates of different gas phases, as well as their relationship to the gas content and star formation rates (SFR) of the host galaxies (\citealt{2013PASJ...65...66S, Adam2015, Paul2018}).

\citet{1985Natur.317...44C} established a canonical steady-state, adiabatic wind model by assuming that supernova (SN) energy is injected thermally and omitting radiative losses. While this model provides foundational scaling relations for mass-loading and velocity, its neglect of radiative cooling often limits its consistency with observations. Subsequent frameworks, such as the momentum-driven model by \citet{2005ApJ...618..569M}, account for the radiative phase of supernova remnants (SNRs), where energy dissipation leads to momentum-dominated evolution. More recently, \citet{2022ApJ...924...82F} introduced a sophisticated model incorporating the multiphase nature of outflows and accounting for mass and energy exchange between cold clouds and the hot wind. Despite these advances, analytical models still struggle to capture the non-linear hydrodynamical complexity inherent in the launching and development of outflows. Specifically, the coupling between feedback and the ISM and the interplay between feedback and star formation activity, are often oversimplified despite their decisive impact on outflow properties.

Numerical simulations provide a powerful tool for investigating stellar-feedback-driven winds across a vast range of spatial scales, spanning from individual giant molecular clouds (GMCs) and SNRs in the ISM, to ``tall-box'' galactic disk models, isolated galaxies, and large-scale cosmological volumes. At GMC scales, simulations have established essential numerical frameworks, such as sink particles for tracking gas collapse, accretion, and star formation (e.g., \citealt{Federrath2010, Gong2012, Howard2016}), and have clarified how cloud properties and early stellar feedback regulate star formation efficiency and cloud destruction (e.g., \citealt{Dale2014, Kim_2018, li2019disruption}). On intermediate scales, high-resolution studies of single SNR evolving in inhomogeneous media have quantified the thermal-to-momentum transition and terminal momentum \citep{Kim_2015}. Furthermore, ``tall-box'' simulations of the multiphase ISM have linked these small-scale processes to the launching and development of galactic winds from gas disks \citep{kim2017three, 2018ApJ...853..173K}. Together, these multi-scale efforts provide the physical foundation for the sub-grid models adopted in larger-volume galaxy simulations.

Based on earlier work \citep[e.g.,][]{Tomisaka1993, 1999ApJ...513..142M}, simulations of M82-like isolated starburst galaxies have made significant progress in producing multiphase winds that allow for direct comparison with observations \citep[e.g.,][]{2008ApJ...674..157C, 2008ApJ...689..153R, melioli2013evolution}. Particularly, recent high-resolution simulations of M82-like galaxies have successfully captured the interplay between star formation, feedback, and the multiphase nature of outflows with great details, including the interactions between cold and hot gas phases \citep[e.g.,][]{schneider2020physical, 2024ApJ...960..100S, schneider2024cgols}. However, starburst events in these simulations are often implemented via simplified assumptions, and the resulting mass outflow rates of the cool phase typically remain lower than those inferred from observations of M82.

Recently, \citet{Wang2026, li2025revisiting} developed a simulation framework that integrates physical modules of star formation, stellar evolution and stellar feedback to mimic the nuclear starburst and multiphase outflow in M82. This approach self-consistently replicates the recent starburst in the nuclear region and reproduces many key observed features of M82's multiphase winds, demonstrating its capacity to capture the essential physics of starburst-driven outflows. While these studies provide a robust foundation, the resolution convergence of the multiphase gas properties remains an area for further optimization.

In contrast, cosmological hydrodynamical simulations, such as EAGLE, IllustrisTNG, and SIMBA \citep{2015MNRAS.446..521S, 2018MNRAS.473.4077P, 2019MNRAS.486.2827D}, utilize sub-grid models to incorporate star formation, stellar feedback and galactic winds, as they lack the resolution to resolve the multiphase interstellar medium (ISM). While these simulations successfully replicate broad statistical properties, including the stellar mass-halo mass relation and the cosmic star formation history, they face significant challenges. Chief among these are the lack of consensus on sub-grid prescriptions of baryonic physics and a heavy reliance on resolution-dependent tuning, which often leads to divergent results as numerical precision increases \citep{2017ARA&A..55...59N, 2023ARA&A..61..473C}.

To improve resolution convergence and better approximate the underlying physical processes in cosmological simulations, significant efforts have been made to refine sub-grid models of star formation and stellar feedback. For instance, \citet{2018MNRAS.480..800H} implemented multiple criteria, requiring gas to be self-gravitating, self-shielding, Jeans unstable, and above a high density threshold, to achieve convergence in galaxy-integrated star formation rates (SFRs). \citet{2024MNRAS.532.3299N} demonstrated that combining the Schmidt law with a gravitational instability criterion replicates the Kennicutt-Schmidt relation while maintaining superior resolution convergence compared to traditional density thresholds and temperature ceilings.

To effectively launch galactic-scale outflows and mitigate numerical overcooling, several supernova (SN) feedback strategies have been developed. Thermal models often employ ``delayed cooling'' \citep{2006MNRAS.373.1074S, 2013MNRAS.429.3068T, 2015MNRAS.454...83W} or high-temperature heating thresholds \citep{2012MNRAS.426..140D} to prevent premature energy dissipation. Alternatively, kinetic \citep{2008A&A...477...79D, 2008MNRAS.387.1431D} or mixed thermal-kinetic \citep{2018MNRAS.473.4077P, 2019MNRAS.486.2827D, 2023MNRAS.523.3709C} models have been widely adopted.

Some mechanical feedback models have gained prominence by utilizing high-resolution simulation results of supernova remnant (SNR) evolution, specifically targeting the terminal momentum at the end of the pressure-driven snowplow phase \citep[e.g.,][]{2014ApJ...788..121K, 2015MNRAS.448.3248G, 2015MNRAS.450..504M}. By injecting the expected momentum into the surrounding interstellar medium, these models better recover the physical solutions of supernova remnants across varying resolutions, leading to improved resolution convergence in star formation and outflow properties \citep{2018MNRAS.478..302S, 2018MNRAS.480..800H, Hu2019, 2020ApJ...900...61K}. 

Despite these advancements, many cosmological sub-grid models still require resolution-dependent calibration of free parameters governing star formation and feedback \citep{2023ARA&A..61..473C,2024MNRAS.532.3417W}. Consequently, predicted baryonic properties, such as the multiphase structure of the CGM and mass loading factors of galactic winds, often diverge from both observational constraints and other high-resolution simulations \citep[e.g.,][]{2020MNRAS.494.3971M,2024MNRAS.531.4560W}. Meanwhile, many idealized galaxy simulations lack the rigorous observational benchmarking necessary to validate their resolution convergence across star formation and outflow properties.

As next-generation cosmological hydrodynamical simulations approach resolutions of a few tens of parsecs, the need for robust, resolution-converged sub-grid models has become increasingly urgent. Achieving this convergence remains a major challenge. As an intermediate step, we use the well-observed M82 starburst galaxy as a controlled testbed to identify and mitigate the dominant numerical divergences inherent in individual-galaxy simulations. Building on the framework developed by \citet{Wang2026} (W26) and \citet{li2025revisiting}, we test, calibrate, and refine star formation and supernova feedback prescriptions across resolutions from several to tens of parsecs. This effort aims to improve the convergence of star formation and outflow properties in simulations within a certain resolution ranges ($4-32\rm\,pc$) of M82-like systems, which are characterized by high gas surface densities and star formation rates. In future work, we will implement these upgrades in zoom-in and cosmological-volume simulations for further validation.

The paper is organized as follows. In Section 2, we detail our numerical methods and simulation setup. Section 3 presents the starburst and outflow properties of our high-resolution fiducial model and examines their convergence across varying resolutions. In Section 4, we compare our modeling strategy with previous work and discuss the critical factors necessary for achieving converged star formation and feedback-driven winds. Finally, we summarize our key findings in Section 5.

\section{Methodology} \label{sec:method}
This section details our numerical setup and physical framework. We first describe the initial conditions for our M82-like galaxy mass model and the key parameters governing the gas disk. We then describe the core baryonic process modules adapted from \citet{Wang2026} and \citet{li2025revisiting}, alongside the specific refinements to star formation and supernova feedback implemented in this work. Lastly, we outline the suite of simulations here.

\subsection{Initial condition} 
The initial conditions and simulation framework are detailed in \citet{Wang2026}, and we provide a brief overview here. We construct a mass model for M82 comprising a gaseous disk, a pre-existing stellar disk, and a dark matter halo. The configuration of the stellar and gas disks follows established methodologies from previous studies \citep[e.g.,][]{Strickland2000, 2008ApJ...674..157C, melioli2013evolution, schneider2020physical}. While earlier models often neglected the dark matter halo \citep[e.g.,][]{Strickland2000, 2008ApJ...674..157C}, its inclusion is essential to reconcile the model with the recent rotation curve observations reported by \citet{Greco2012}.

We adopt a \citet{Miyamoto1974} profile for the stellar disk, assuming a mass of $1.5 \times 10^9\,M_\odot$ with a scale length and height of $850$ pc and $250$ pc, respectively. Although M82 contains a bulge of $\sim 5.0 \times 10^8\,M_\odot$ within the central $1$ kpc, a significant fraction of this mass likely formed during the recent starburst. Given our focus on the past $20$–$30$ Myr of starburst activity, we incorporate the current bulge mass into the initial gas component following \citet{Wang2026}. The dark matter halo is represented by a core-NFW profile with a mass of $M_{\rm dmh} \approx 8.5 \times 10^{10}\,M_\odot$ and a virial radius of $R_{\rm vir} = 53$ kpc. Furthermore, the gas disk is modeled using a triple Miyamoto-Nagai profile \citep[e.g.,][]{Smith2015}, with a total mass of $3.5 \times 10^9\,M_\odot$ and a half-mass radius of $r_{\rm hm} = 1$ kpc. The parameters for each component are carefully calibrated to ensure the resulting total rotation curve reproduces the observations from \citet{2012ApJ...757...24G}.

The initial density distributions for the gaseous disk, the bulge (initialized as gas), and the ambient hot halo are defined by
\begin{equation}
\begin{split}
    &\rho_{d}(r_{ax},z) =  \rho_{d,0}  \times \\
    & \rm{exp}\left[ -\frac{\Phi_{tot}(r_{ax},z)-e_{d}^{2}\Phi_{tot}(r_{ax},0)-(1-e_{d}^{2})\Phi_{tot}(0,0)}{c_{s,d}^{2}} \right]
\end{split}
\end{equation}

\begin{equation}
\begin{split}
    &\rho_{b}(r_{ax},z) =  \rho_{b,0}  \times \\
    & \rm{exp}\left[ -\frac{\Phi_{tot}(r_{ax},z)-e_{b}^{2}\Phi_{tot}(r_{ax},0)-(1-e_{b}^{2})\Phi_{tot}(0,0)}{c_{s,b}^{2}} \right]
\end{split}
\end{equation}

\begin{equation}
\begin{split}
    &\rho_{h}(r_{ax},z) =  \rho_{h,0}   \times \\
    & \rm{exp}\left[ -\frac{\Phi_{tot}(r_{ax},z)-e_{h}^{2}\Phi_{tot}(r_{ax},0)-(1-e_{h}^{2})\Phi_{tot}(0,0)}{c_{s,h}^{2}} \right] 
\end{split}
\end{equation}
Here, $r_{\rm ax} = \sqrt{x^2 + y^2}$ denotes the cylindrical radius and $z$ represents the vertical coordinate. The variables $\rho_{d,0}$, $\rho_{b,0}$, and $\rho_{h,0}$ represent the initial central densities of the disk, bulge, and halo components, respectively, while $\Phi_{\rm tot}$ is the total gravitational potential from the stellar disk, gas disk, and dark matter halo. The rotational support factors are given by, 
\begin{equation}
    e_{d} = e_{rot,d} \  \rm{exp}(-z/z_{rot}),
\end{equation}

\begin{equation}
    e_{b} = e_{rot,b} \  \rm{exp}(-z/z_{rot}),
\end{equation}

\begin{equation}
    e_{h} = e_{rot,h} \  \rm{exp}(-z/z_{rot}).
\end{equation}
The parameters for these components are summarized in Table \ref{tab:disk&halo}. Within our $8$ kpc cubic simulation volume, the initial total gas mass is $M_{\rm gas,tot} = 3.2 \times 10^9\,M_\odot$. Approximately $95\%$ of this mass is concentrated near the midplane, within $|z| < 400$ pc. 

\begin{table}[ht]
\centering
\caption{Initial parameters for the gas disk, bulge, and halo components. Central density ($\rho_0$), rotation factor ($e_{\rm rot}$), and effective sound speed ($c_s$) are listed. The bulge is initialized as a gas component (see text).}
\begin{tabular}{lccc}
\hline
\hline
Component & $\rho_{0}$ (cm$^{-3}$) & $e_{\rm rot}$ & $c_{s}$ (cm s$^{-1}$) \\
\hline
Gas disk     & 320                  & 0.92          & $3.6 \times 10^{6}$ \\
Bulge        & 70                   & 0.50          & $5.7 \times 10^{6}$ \\
Hot gas halo & $2.0 \times 10^{-3}$ & 0.25          & $3.0 \times 10^{7}$ \\
\hline
\end{tabular}
\label{tab:disk&halo}
\end{table}

\subsection{Overall view of Baryonic Physics}
The physical modules incorporated in our simulations, including radiative cooling and heating, star formation, stellar winds, radiation pressure, photoionization feedback, core-collapse supernovae (CCSNe), gaseous mass return from sink particles, and gravity, are based on the framework established in \citet{Wang2026} and \citet{li2025revisiting}. Below, we summarize these modules and describe the refinements implemented to enhance resolution convergence.

The gravitational potential from the dark matter and stellar components is treated as a static background, while gas self-gravity and gas--sink/star particle gravitational interactions are computed at each timestep. We track the gas dynamics and implement self-consistent star formation using a sink particle method based on \citep{Federrath2010}. Stellar feedback, encompassing stellar winds, radiation pressure, photoionization feedback, and CCSNe is modeled following several distinct prescriptions. Stellar winds are implemented using the metallicity-dependent model of \citet{Jorick2021}. CCSN feedback utilizes the variable progenitor model from \citet{Sukhbold2015}, implemented via the compound kinetic-thermal feedback scheme described in \citet{kim2017three}. Additionally, we account for the cumulative effects of clustered supernovae \citep[e.g.,][]{Gentry17}.

Radiative cooling and heating are implemented via the Grackle library \footnote{https://grackle.readthedocs.io} \citep{Smith2016}, with a temperature floor of $300$ K. 
Following \citet{Wang2026}, we include the built-in photoelectric heating module of GRACKLE for heating rate calculation, which is based on equation (1) in \citet{Wolfire1995}.
Photoionization feedback from young star clusters (star particles in our simulations) is modeled by a Stromgren-type approximation (e.g. \citet{2021MNRAS.506.3882S}) based on equation (23) in \citet{Wang2026}. Radiation pressure, driven by the local UV flux from star particles, is also implemented following the methodology in \citet{Wang2026}. To model the disruption of giant GMCs, we account for the mass return from sink/star particles back to the ISM using the prescriptions defined in Equations (5) and (6) of \citet{li2025revisiting}. Additionally, we implement a passive scalar variable to track the enrichment and transport of metals throughout the simulation.

In this work, we refine two key modules from \citet{Wang2026} and \citet{li2025revisiting} to improve resolution convergence: the star formation prescription and the coupling of CCSN feedback to the ISM. These upgrades are designed to produce consistent results for star formation, feedback efficiency, and outflow properties in simulations of an M82-like galaxy across resolutions ranging from several to tens of parsecs. The inherited and revised prescriptions are summarized in Table~2. Section 2.3 and 2.4 describe the modifications to star formation and coupling of CCSN feedback, respectively; unchanged procedures are described in W26 and L25.

We define the model configurations used throughout this paper before describing the individual modifications. The reference model, OS-FNG-mw, adopts the original W26/L25 star formation prescription, with a sink control volume that scales with grid size and a fixed star formation efficiency per free-fall time. It also uses the fixed-number-of-cells SN coupling region and mass-weighted deposition scheme of L25. Our fiducial upgraded model, NS-RCO-gs, adopts the revised star formation prescription described in Section~2.3, together with an SN coupling region tied to the physical cooling scale and a Gaussian-weighted deposition kernel described in Section~2.4.

We further perform simulations to isolate the effects of individual modifications. HS-FNG-mw replaces the resolution-scaled sink control volume with a nearly fixed physical-scale volume, while HS2-FNG-mw tests the resolution- and density-dependent star formation efficiency with the original control-volume geometry. OS-RCO-mw applies the revised SN coupling scale to the original star formation prescription. Finally, NS-FNG-mw and NS-RCO-mw isolate the effects of the revised CCSN coupling prescriptions with the updated star formation model. The complete model configurations are listed in Table~\ref{tab:sim_list}.

\begin{figure*}[!htbp]
    \centering
    \includegraphics[width=0.8\textwidth]{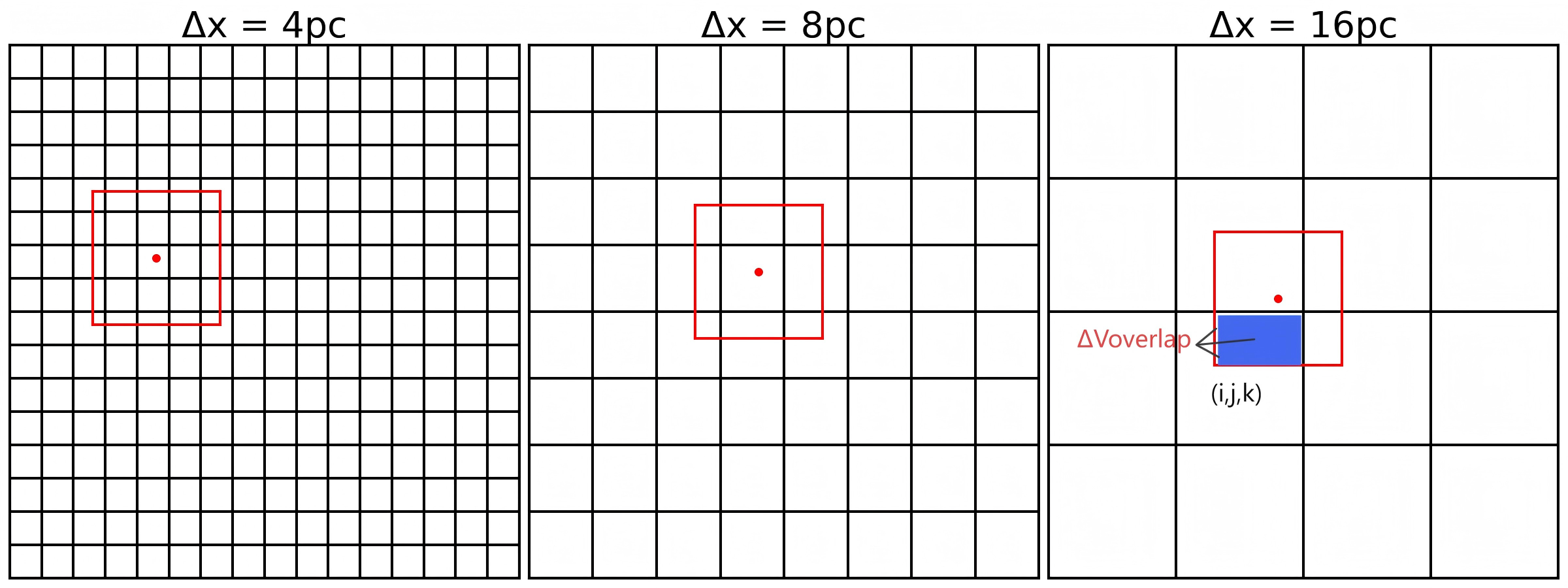}
    \caption{Two-dimensional schematic of a fixed-scale accretion control volume, $V_{\rm cv} = (16\,\text{pc})^3$, across varying numerical resolutions. Panels from left to right correspond to grid resolutions of $\Delta x = 4, 8,$ and $16\,\text{pc}$, respectively. The red dot denotes the sink particle, while the red outline defines the physical extent of the control volume. The blue shaded regions illustrate the overlapping volume between an individual grid cell $(i, j, k)$ and the sink particle’s control volume}
    
    \label{fig:accrete}
\end{figure*}

\subsection{Star Formation Refinements}
\label{sec:sf_model}

As described in \citet{Wang2026}, sink particles are created in collapsing regions satisfying the Jeans-density, converging-flow, non-overlap, and gravitational-potential-minimum criteria. Once formed, they accrete gas within their control volume and convert it into stars at a prescribed rate. Thus, sink particles represent bound, collapsing gas clumps rather than resolved individual stars. When the stellar mass in a sink reaches a specified threshold, $M_{\rm th}$, the sink is converted into a star particle and gas accretion ceases. 

We adopt $M_{\rm th}=10^4\,M_\odot$ for $\Delta x\leq16$ pc and $M_{\rm th}=2\times10^4\,M_\odot$ for $\Delta x=32$ pc, which differ from \citet{Wang2026} moderately. The typical accretion timescale, spanning from sink instantiation to star particle formation, is $\sim 1$ Myr. We track the total mass of the sink/star particle containing both gas and stellar content as $M_{\rm sink,tot}(t)=M_{\rm sink,gas}(t)+M_{\rm sink, \star}(t)$. The treatments of particle dynamics, gas accretion, mass, metal and momentum conservation,  merging of nearby sinks, sink-star conversion, and gravitational interactions are detailed in \citet{Wang2026}.

Here, we describe the main modifications to the sink-particle and star formation prescriptions in this work. The first modification concerns the sink control volume and accretion prescription. We adopt a nearly fixed physical-scale cubic control volume, instead of the spherical volume with radius $1.5\Delta x$ used in \citet{Wang2026}. For $\Delta x\leq16$ pc, we set $V_{\rm cv}=(16\,{\rm pc})^3$, while for the coarsest resolution, $\Delta x=32$ pc, we use $V_{\rm cv}=(32\,{\rm pc})^3$ to ensure that it contains at least one grid cell. The same physical control volume is used for overlap checks when creating sink particle.

At each time step, sink particles accrete gas from grid cells intersecting their control volumes, considering only cells with $\rho>\rho_{\rm th}$. Following \citet{Wang2026}, we define the sink-formation and accretion density threshold as

\begin{equation}
    \rho_{\rm th} = \frac{\pi}{16} \frac{c_{s}^{2}}{G\Delta x^{2}},
\end{equation}
where $c_s$ is the local sound speed and $\Delta x$ is the cell size. This corresponds to adopting the Jeans criterion $\lambda_J>4\Delta x$ following \citet{1997ApJ...489L.179T}.

In this work, only the fraction of each gas cell overlapping the control volume contributes to sink accretion,
\begin{equation}
    \Delta M_{\rm acc}(i,j,k)
    =
    \left[\rho(i,j,k)-\rho_{\rm th}(i,j,k)\right]
    \Delta V_{\rm overlap}(i,j,k).
\end{equation}
This differs from \citet{Wang2026}, who use the full cell volume $\Delta V$. Figure~\ref{fig:accrete} illustrates the control and overlapping volumes at different resolutions.

\begin{figure*}[!htbp]
    \centering
    \includegraphics[width=0.75\textwidth]{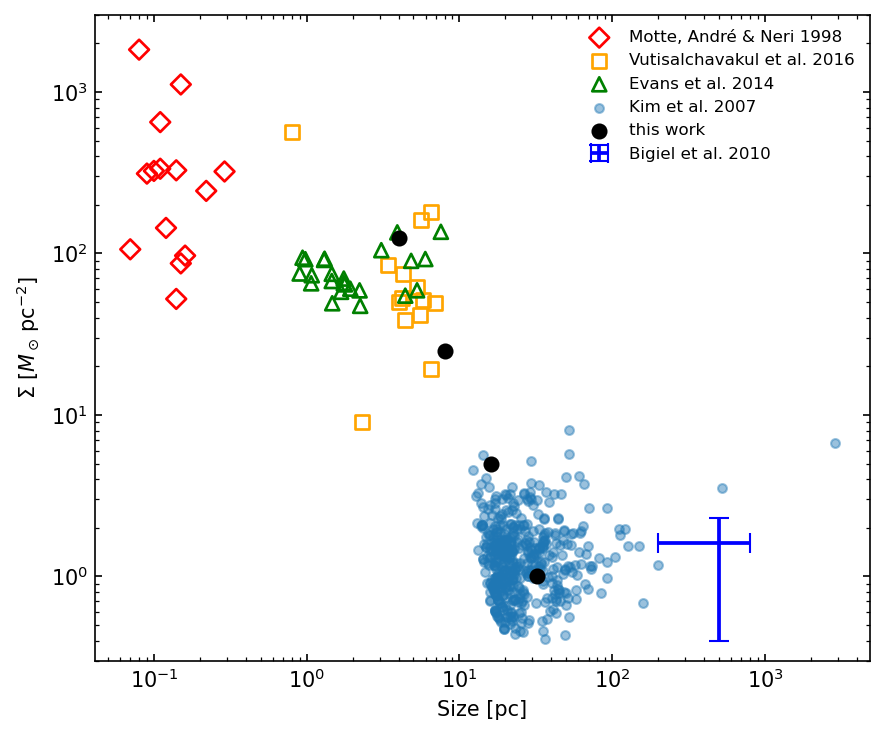}
    \caption{The surface-density of molecular cloud (\citealt{1998A&A...336..150M, 2014ApJ...782..114E, 2016ApJ...831...73V}), HI cloud \citep{2007ApJS..171..419K} and outer disk HI gas \citep{2010AJ....140.1194B} at corresponding size or spatial scale in observations and $\Sigma_{0}$ adopted in our simulations (black dot).}
    
    \label{fig:sigma_size}
\end{figure*}

Note that, when the control volumes of two sink particles overlap, their treatment depends on their combined stellar mass relative to the conversion threshold $M_{\rm th}$ \citep{Wang2026}. If the combined mass is below $M_{\rm th}$, the sinks merge into one particle, with stellar, gas, and metal masses summed and the position and velocity set to mass-weighted averages. If it exceeds $M_{\rm th}$, they do not merge; instead, sufficient stellar mass is transferred from the less massive sink to the more massive one for the latter to reach $M_{\rm th}$ and convert into a star particle. The corresponding gas and metal masses are transferred proportionally, while the donor remains a sink. Linear momenta are updated throughout the transfer to conserve total momentum.

The second main modification concerns star formation rate and efficiency for each sink particle, especially the star-formation efficiency (SFE) per local free-fall time. At each time step, the stellar mass growth rate (i.e.,
the gas-to-star conversion rate) for each sink particle is 
\begin{equation}
    \label{eq:mass_convert}
    \dot M_{\rm sink, \star} (t) = \frac{M_{\rm sink,gas}(t)+M(i,j,k)}{t_{\rm sf,clump}},
\end{equation}
where $M_{\rm sink, gas}(t)$ is the gas mass stored in this sink particle after accretion, $M(i,j,k)=\rho(i,j,k)\Delta V(i,j,k)$ is the gas mass of the grid cell containing the sink particle, and $t_{\rm sf,clump}$ is the local star formation timescale at the clump scale, defined as:

\begin{equation} 
\label{eq:tsf} 
t_{\rm sf,clump} = \frac{t_{\rm ff,clump}}{\epsilon_{\rm ff,clump}} = \frac{1}{\epsilon_{\rm ff,clump}} \sqrt{\frac{3 \pi}{32 G \rho_{\rm clump}}}, \end{equation}
where the clump density, $\rho_{\rm clump}$, was defined as:

\begin{equation}
\label{eq:rho_clump}
\rho_{\rm clump} = \frac{M_{\rm sink,tot}(t) + M(i,j,k)}{\Delta x^{3}}, 
\end{equation}

and $\epsilon_{\rm ff,clump}$ is the star formation efficiency per free-fall time at the clump scale.

Note that \citet{Wang2026} does not include $M(i,j,k)$ when defining $t_{\rm sf,clump}$ and $\rho_{\rm clump}$, which may underestimate the local star formation efficiency. In our simulations, sink particles form in dense, cold gas, with typical temperatures ranging from the temperature floor ($\sim300$ K) to a few $10^3$ K.

While \citet{Wang2026} adopts a fixed $\epsilon_{\rm ff,clump}=0.2$, we introduce a density- and resolution-dependent prescription to recover the empirical star formation law $\Sigma_{\rm SFR}\propto\Sigma_{\rm gas}^{N}$ (e.g., \citealt{1998ApJ...498..541K}) at the sink/clump scale and improve resolution convergence (see Appendix A):

\begin{equation}
\label{eq:sfr_relation_eff}
\epsilon_{\rm ff,clump} = \epsilon_{\rm 0,ff} \left( \frac{\Sigma_{0}}{\Sigma_{\rm clump}} \right)^{1.5-N},
\end{equation}
where $\Sigma_{\rm clump}=\rho_{\rm clump} \Delta x$, and $\epsilon_{\rm 0,ff}$, $N$, and $\Sigma_0$ are specified below.

We adopt $\epsilon_{0,\rm ff}=0.1$, approximately midway between the observed $\epsilon_{\rm ff}\sim3\%$ on molecular-cloud scales ($\sim4$--$28$ pc) and $\sim20\%$ on dense clump/core scales ($\sim1.5$ pc) \citep{2025MNRAS.540.2377R}. The classic Kennicutt--Schmidt law gives $N\approx1.5$ on galactic scales, while observations from kpc to sub-pc scales find $1\lesssim N\lesssim3$ \citep[e.g.,][]{2008AJ....136.2846B, 2009ApJS..181..321E, 2010ApJ...723.1019H, 2025MNRAS.540.2377R}. We adopt $N=1.8$, motivated by theoretical models and simulations suggesting steeper star-formation laws at higher cold-gas densities \citep{2010MNRAS.407.2091G}, as expected for the M82-like system studied here.

To improve consistency across resolutions, we scale $\Sigma_0$ with $\Delta x$, adopting $\Sigma_0=125$, $25$, $5$, and $1.0\,M_\odot\,\mathrm{pc}^{-2}$ for $\Delta x=4$, $8$, $16$, and $32$ pc, respectively. These values approximately match the observed surface densities of interstellar gas structures on comparable scales, including molecular clouds \citep{1998A&A...336..150M, 2014ApJ...782..114E, 2016ApJ...831...73V}, HI clouds \citep{2007ApJS..171..419K}, and outer-disk HI gas \citep{2010AJ....140.1194B} (Figure~\ref{fig:sigma_size}). Although this scaling is not derived from a fundamental physical model, it provides an observationally motivated prescription and, as shown in Section~3, is important for achieving resolution convergence in the predicted star-formation properties.

On the other hand, we also compared the characteristic density $\rho_0$ associated with $\Sigma_0$ to the threshold density $\rho_{\rm th}$ at different resolutions in one group of simulation. We find that $\rho_0$ is comparable to $\rho_{\rm th}$ at $\Delta x=4\,\mathrm{pc}$, but becomes progressively lower than $\rho_{\rm th}$ as the resolution is coarsened beyond $8\,\mathrm{pc}$. The discrepancy increases systematically with decreasing resolution. This behavior suggests that $\Sigma_0$ and $\rho_{\rm th}$ are not redundant parameters. While $\rho_{\rm th}$ is set by the sink formation criterion, $\Sigma_0$ serves as a calibration scale for the resolution- and density-dependent star formation efficiency per local free-fall time.

\newlength{\CompareRowPad}
\setlength{\CompareRowPad}{7pt}
%

\newcommand{\TblCellL}[3]{%
  \parbox[c][\dimexpr#1+\CompareRowPad\relax][c]{#2}{%
    \raggedright
    #3\par
  }%
}

%

\newcommand{\TblCellC}[3]{%
  \parbox[c][\dimexpr#1+\CompareRowPad\relax][c]{#2}{%
    \centering
    #3\par
  }%
}

%

\newcommand{\CompareRow}[5]{%
  \TblCellL{#1}{3.5cm}{#2}
  &
  \TblCellC{#1}{5.0cm}{#3}
  &
  \TblCellC{#1}{5.0cm}{#4}
  &
  \TblCellL{#1}{3.8cm}{#5}
  \\
}

\begin{table*}[htbp]
\centering

\caption{Comparison of the inherited star formation and SN feedback coupling prescriptions from W26/L25 with the modified prescriptions adopted in the fiducial model of this work.}
\label{tab:compare_model}

\small

\renewcommand{\arraystretch}{1.0}

\setlength{\tabcolsep}{3pt}

\begin{tabular}{@{}cccc@{}}
\toprule


\TblCellL{0.70cm}{3.5cm}{%
\textbf{Model Ingredient}
}
&
\TblCellC{0.70cm}{5.0cm}{%
\textbf{W26/L25}
}
&
\TblCellC{0.70cm}{5.0cm}{%
\textbf{This work}
}
&
\TblCellC{0.70cm}{3.8cm}{%
\textbf{Reference}
}
\\[1pt]

\midrule


\multicolumn{4}{@{}l@{}}
{\textbf{Sink particle and star formation}}
\\[3pt]


\CompareRow{1.55cm}
{%
Sink creation criteria
}
{%
Jeans-density, non-overlap,
converging flow, gravitational
potential minimum
}
{%
Unchanged
}
{%
\citet{Federrath2010};
\citet{Wang2026}
}


\CompareRow{0.68cm}
{%
Sink particle mass
}
{%
$M_{\rm sink,tot}
=
M_{\rm sink,gas}
+
M_{\rm sink,\star}$
}
{%
Unchanged
}
{%
\citet{Wang2026}
}


\CompareRow{0.68cm}
{%
Sink particle metallicity
}
{%
$Z_{\rm sink}
=
M_{\rm sink,metal}/M_{\rm sink,tot}$
}
{%
Unchanged
}
{%
\citet{Wang2026}
}


\CompareRow{1.45cm}
{%
Control-volume geometry
}
{%
Spherical:
\\[2pt]
$r_{\rm cv}=1.5\Delta x$
}
{%
Cubic:
\\[2pt]
$V_{\rm cv}=(16\,{\rm pc})^3$
for $\Delta x\le16\,{\rm pc}$;
\\
$V_{\rm cv}=(32\,{\rm pc})^3$
for $\Delta x=32\,{\rm pc}$
}
{%
This paper
}


\CompareRow{0.95cm}
{%
Gas accretion onto sink
from a cell
}
{%
$\Delta M_{\rm acc}
=
(\rho-\rho_{\rm th})
\Delta V_{\rm cell}$
}
{%
$\Delta M_{\rm acc}
=
(\rho-\rho_{\rm th})
\Delta V_{\rm overlap}$
}
{%
\citet{Federrath2010}
}


\CompareRow{1.20cm}
{%
Clump density
}
{%
Defined from sink particle mass:
\\[3pt]
$\rho_{\rm clump}
=
M_{\rm sink,tot}/\Delta x^3$
}
{%
Host-cell gas mass is included:
\\[3pt]
$\rho_{\rm clump}
=
(M_{\rm sink,tot}+M_{\rm host})/\Delta x^3$
}
{%
\citet{Wang2026}
}


\CompareRow{1.35cm}
{%
Star-formation efficiency
per free fall time
}
{%
$\epsilon_{\rm ff}=0.2$
}
{%
Density- and resolution-dependent:
\\[3pt]
$\epsilon_{\rm ff}
=
\epsilon_{0,\rm ff}
\left(
\frac{\Sigma_0}{\Sigma_{\rm clump}}
\right)^{1.5-N}$
}
{%
This paper
}


\CompareRow{1.00cm}
{%
Sink stellar-mass growth
}
{%
$\dot M_{\rm sink,\star}
=
\epsilon_{\rm ff}
M_{\rm sink,gas}
/t_{\rm ff,clump}$
}
{%
$\dot M_{\rm sink,\star}
=
\epsilon_{\rm ff}
(M_{\rm sink,gas}+M_{\rm host})
/t_{\rm ff,clump}$
}
{%
\citet{Wang2026}
}


\CompareRow{1.60cm}
{%
Sink-to-star conversion
threshold
}
{%
$M_{\rm th}\propto\Delta x$:
\\[3pt]
$5\times10^{4}-10^{5}\,M_\odot$
\\
in W26 for $\Delta x=4$--$8\,{\rm pc}$
}
{%
$M_{\rm th}=10^{4}\,M_\odot$
\\
for $\Delta x\le16\,{\rm pc}$;
\\[2pt]
$M_{\rm th}=2\times10^{4}\,M_\odot$
\\
for $\Delta x=32\,{\rm pc}$
}
{%
\citet{Wang2026}
}



\midrule

\multicolumn{4}{@{}l@{}}
{\textbf{Core-collapse supernova feedback}}
\\[3pt]


\CompareRow{1.35cm}
{%
SN coupling radius
}
{%
Fixed cell number:
\\[3pt]
$r_{\rm snr}=3\Delta x$ 
}
{%
$\sim R_{cool}$:
\\[3pt]
$r_{\rm snr}=4,3,2,1\Delta x$
\\
for $\Delta x=4,8,16,32\,{\rm pc}$
}
{%
\citet{li2025revisiting}
}


\CompareRow{2.20cm}
{%
Energy/momentum-mode switch
}
{%
$R_M>1$:
final momentum injection;
\\[4pt]
$0.027<R_M<1$:
28\% kinetic and
72\% thermal energy;
\\[4pt]
$R_M<0.027$:
100\% kinetic energy
}
{%
$R_M>1$:
final momentum injection;
\\[5pt]
$R_M<1$:
50\% kinetic and
50\% thermal energy
}
{%
\citet{kim2017three}
}


\CompareRow{1.5cm}
{%
Deposition kernel
}
{%
Mass-weighted injection scheme
}
{%
Gaussian kernel with kernel width:
\\[3pt]
$\sigma=4,3,2,1\Delta x$
\\
for $\Delta x=4,8,16,32\,{\rm pc}$
}
{%
This paper
}


\bottomrule

\end{tabular}

\end{table*}

\subsection{Core-Collapse Supernova Feedback and Refinement}
\label{sec:CCSN}

Core-collapse supernovae (CCSNe) are the primary drivers of galactic outflows in our simulations. Following \citet{li2025revisiting}, we adopt a CCSNe feedback model based on the compound prescription of \citet{kim2017three}, in which the feedback coupling depends on whether the local Sedov--Taylor phase is numerically resolved, rather than being prescribed through a fixed thermal-energy or momentum injection scheme. Within this framework, the key parameter is the ratio
\begin{equation}
\label{eq:Rm}
R_{M} = \frac{M_{snr}}{M_{\rm sh}}
\end{equation}
where \(M_{\rm snr}\) is the total gas mass within the feedback coupling region, including the SN ejecta, and \(M_{\rm sh}\) is the shell-formation mass of the supernova remnant. When \(R_M > 1\), the Sedov--Taylor phase is unresolved or the remnant is already radiative, so feedback is injected as terminal radial momentum. When \(R_M < 1\), the remnant is sufficiently resolved and the SN energy is deposited as a mixed kinetic--thermal input. The transition between energy-conserving and momentum-conserving evolution is therefore handled explicitly through the \(R_M\)-dependent switching in the compound feedback prescription.

Specifically, for SN events with $R_{M}>1$, we inject the terminal radial momentum
\begin{equation}
\label{eq:psnr}
p_{snr}=2.8\times10^{5}\, M_{\odot}\,\mathrm{km\,s^{-1}}(\frac{n_{H}}{\mathrm{cm^{-3}}})^{-0.17}f_{\rm boost}
\end{equation}
where $f_{\rm boost}$ accounts for the momentum enhancement expected from clustered supernovae following \citet{Gentry17}. The enhanced terminal momentum in \citet{Gentry17} originates from an extended energy-conserving phase maintained by a sharp contact discontinuity, which is only realized in idealized spherical simulations. In more realistic, inhomogeneous media, radiative shell instabilities and turbulent mixing promote efficient cooling and significantly shorten the energy-conserving stage (\citealt{2017ApJ...834...25K, 2020MNRAS.492.1243G, 2021ApJ...914...89L, 2021ApJ...914...90L, 2024ApJ...970...18L}).

Here, $f_{\rm boost}$ should be interpreted as a sub-grid correction for overlapping and temporally clustered supernovae, rather than implying that the full momentum enhancement of idealized spherical remnants is realized in our stratified, inhomogeneous ISM. Accordingly, we apply $f_{\rm boost}=10$ following the \citet{li2025revisiting} only within the compound framework described above, while ensuring the conservation of the total energy budget from single SN event. For $R_{M}<1$, we divide the total SN energy equally between kinetic and thermal components, which differs moderately from the prescription used in \citet{li2025revisiting}.

Compared to \citet{li2025revisiting}, our refinement in this work concerns the spatial scale of the coupling region and the weighting scheme used to distribute SN ejecta within it. In \citet{li2025revisiting}, the feedback radius was fixed at $r_{\rm snr} = 3\Delta x$, and the ejecta were distributed using a mass-weighted injection scheme. Under this prescription, the physical coupling volume increases linearly with the grid size, so that at coarse resolution the ejecta are deposited over an artificially large volume, effectively increasing $M_{\rm snr}$. This in turn causes a larger fraction of SN events to satisfy the $R_M > 1$ condition, leading to enhanced numerical overcooling.

For reference, under homogeneous and isotropic conditions, the cooling radius of a single supernova remnant is given by \citet{Kim_2015}:
\begin{equation}
\label{eq:Lcoool}
R_{\rm cool} = 22.6\,\text{pc } E_{51}^{0.29} (n_{0,\text{ISM}} / \text{cm}^{-3})^{-0.42}
\end{equation}
where $E_{51}$ is the explosion energy in unit of $10^{51}\,\text{erg}$, and $n_{0,\rm ISM}$ is the ambient number density.

From Equation~\ref{eq:Lcoool}, one possible solution to avoid numerical overcooling is to reduce the spatial scale of the coupling region such that $r_{\rm snr} < R_{\rm cool}$. However, for ambient densities $n_{0,\rm ISM} \gtrsim 1\,\mathrm{cm^{-3}}$, the cooling radius quickly drops below $\sim 10\,\mathrm{pc}$, making this ad hoc adjustment of $r_{\rm snr}$ ineffective in simulations with coarse resolution. Nevertheless, this issue is mitigated by supernova clustering: early explosions evacuate the surrounding medium, allowing subsequent SNe to occur in lower-density cavities where $R_{\rm cool}$ can exceed $\sim 20\,\mathrm{pc}$ and can therefore be adequately resolved even at relatively modest numerical resolution.

For a canonical explosion energy of $10^{51}\,\mathrm{erg}$ and ISM conditions representative of our galactic model, specifically a disk, averaged post-SN number density of $\bar{n} \sim 1{-}10\,\mathrm{cm^{-3}}$, Equation~\ref{eq:Lcoool} implies a characteristic cooling radius of $R_{\rm cool} \sim 20\,\mathrm{pc}$. In this work, we choose $r_{\rm snr}$ to track this physical scale as closely as possible on the discrete mesh. For the resolutions considered here, this yields $r_{\rm snr} = 4\Delta x, 3\Delta x, 2\Delta x$, and $\Delta x$ for $\Delta x = 4, 8, 16$, and $32\,\mathrm{pc}$, respectively. This resolution-dependent choice is designed to ensure that the compound $R_M$-based feedback prescription operates as consistently as possible across different resolutions (see Section 4.2), thereby improving convergence of the resulting outflow properties.

\citet{li2025revisiting} adopted a mass-weighted injection scheme to avoid depositing excessive energy into very low-density cells, which can otherwise generate unphysically high temperatures and prohibitively small Courant time-steps. However, in stratified disk simulations, this approach would inject a considerable fraction of ejecta momentum horizontally toward the dense midplane, rather than isotropically, a known limitation in previous study \citep{Hopkins18}. To mitigate this effect, we instead adopt a Gaussian-weighted kernel as our fiducial scheme. This provides a smoother and less density-biased distribution of mass, momentum, and energy, while preserving the same $R_M$-dependent compound feedback framework.

Detailed implementation of supernova feedback is given in Appendix \ref{sec:sn_feedback}.

\subsection{Simulations}
We performed a suite of 3D hydrodynamical simulations using the ATHENA++ code \citep{2020ApJS..249....4S} including self-gravity. Our primary objective is to evaluate resolution convergence using the updated star formation and CCSNe feedback modules described above; other physics, such as stellar winds and gas return, follow the default parameters in \citet{li2025revisiting}. The simulations utilize an Static Mesh Refinement (SMR) structure within an 8 kpc cubic box. The highest refinement level is concentrated in the gas disk center ($R < 1000$ pc, $|z| < 120$ pc), where the cell size defines the simulation's nominal resolution, $\Delta x$. Outside this region, the grid size increases by successive factors of two. We refer readers to Appendix B of \citet{Wang2026} for an illustration of the mesh refinement strategy.
We conducted four primary runs with $\Delta x = 4, 8, 16,$ and $32$ pc to rigorously test the convergence of our new prescriptions.

Our simulations are designed to replicate the recent starburst and multiphase galactic winds observed in M82 over a duration of $50\,\rm{Myr}$. We perform seven distinct groups of simulations, each comprising four runs with nominal resolutions of $\Delta x = 4, 8, 16,$ and $32\,\rm{pc}$. These groups are categorized based on their star formation (SF) prescriptions and supernova (SN) feedback injection schemes, the complete parameter space and group designations are summarized in Table ~\ref{tab:sim_list}.

\begin{deluxetable*}{ccccccccc}
\tablecaption{Summary of the simulation suite. The seven simulation groups are named based on their star formation and feedback prescriptions. Star Formation: "OS" refers to the original scheme from \citet{li2025revisiting} with a resolution-dependent control volume ($1.5 \Delta x$) and a constant efficiency ($\epsilon_{\rm ff,clump} = 20\%$); "NS" denotes the new scheme with a fixed physical-scale control volume and resolution/density-dependent efficiency; "HS" utilizes a fixed physical-scale control volume with $\epsilon_{\rm ff,clump} = 20\%$; "HS2" adopts the same spherical control volume ($r = 1.5\Delta x$) as ``OS'' but a resolution/density-dependent efficiency; $\Sigma_0$,  $\epsilon_{\rm 0,ff}$ and the power-law index $N$ regulate the star formation rate's sensitivity to the local gas surface density. SN Feedback: "FNG" and "RCO" indicate feedback deposition radii fixed to a multiple of the grid size ($3 \Delta x$) or approximate to the physical cooling radius ($R_{\rm cool}$), respectively. The suffixes "mw" and "gs" distinguish between mass-weighted and Gaussian-weighted injection kernels. The "resolution" column lists the nominal cell size ($\Delta x$) in the refined central region. Note that for all runs, the deposition radius for stellar winds and returning gas is fixed at $2 \Delta x$, and this common parameter is omitted from the group names. \label{tab:sim_list}}
\tablehead{
\colhead{Simulations groups} & \colhead{resolution ($\Delta x$/pc)} & \colhead{Control volume} & 
\colhead{$\rm M_{th}$ ($\rm {M_{\odot}}$)} & \colhead{$\Sigma_{0}$ ($\rm {M_{\odot} \,pc^{-2}}$)} & 
\colhead{$\epsilon_{\mathrm{0,ff}}$} & \colhead{$N$} & \colhead{$r_{snr}$}
}
\startdata
\multirow{4}{*}{OS-FNG-mw} 
 & $4 $ & $r_{cv}=1.5\Delta x$ & $1.0\times10^{4}$ & \textbackslash & \textbackslash & \textbackslash & $3\Delta x$ \\
 & $8 $ & $r_{cv}=1.5\Delta x$ & $2.0\times10^{4}$ & \textbackslash & \textbackslash & \textbackslash & $3\Delta x$ \\
 & $16 $ & $r_{cv}=1.5\Delta x$ & $4.0\times10^{4}$ & \textbackslash & \textbackslash & \textbackslash & $3\Delta x$ \\
 & $32 $ & $r_{cv}=1.5\Delta x$ & $8.0\times10^{4}$ & \textbackslash & \textbackslash & \textbackslash & $3\Delta x$ \\
 \tableline
\multirow{4}{*}{NS-RCO-gs} 
 & $4 $ & $V_{cv}=(16\rm\,pc)^{3}$ & $1.0\times10^{4}$ & 125 & 0.1 & 1.8 & $4\Delta x$ \\
 & $8 $ & $V_{cv}=(16\rm\,pc)^{3}$ & $1.0\times10^{4}$ & 25 & 0.1 & 1.8 & $3\Delta x$ \\
 & $16 $ & $V_{cv}=(16\rm\,pc)^{3}$ & $1.0\times10^{4}$ & 5 & 0.1 & 1.8 & $2\Delta x$ \\
 & $32 $ & $V_{cv}=(32\rm\,pc)^{3}$ & $2.0\times10^{4}$ & 1 & 0.1 & 1.8 & $1\Delta x$ \\
 \tableline
\multirow{4}{*}{HS-FNG-mw} 
 & $4 $ & $V_{cv}=(16\rm\,pc)^{3}$ & $1.0\times10^{4}$ & \textbackslash & \textbackslash & \textbackslash & $3\Delta x$ \\
 & $8 $ & $V_{cv}=(16\rm\,pc)^{3}$ & $1.0\times10^{4}$ & \textbackslash & \textbackslash & \textbackslash & $3\Delta x$ \\
 & $16 $ & $V_{cv}=(16\rm\,pc)^{3}$ & $1.0\times10^{4}$ & \textbackslash & \textbackslash & \textbackslash & $3\Delta x$ \\
 & $32 $ & $V_{cv}=(32\rm\,pc)^{3}$ & $2.0\times10^{4}$ & \textbackslash & \textbackslash & \textbackslash & $3\Delta x$ \\
 \tableline
\multirow{4}{*}{HS2-FNG-mw} 
 & $4 $ & $r_{cv}=1.5\Delta x$ & $1.0\times10^{4}$ & 125 & 0.1 & 1.8 & $3\Delta x$ \\
 & $8 $ & $r_{cv}=1.5\Delta x$ & $2.0\times10^{4}$ & 25 & 0.1 & 1.8 & $3\Delta x$ \\
 & $16 $ & $r_{cv}=1.5\Delta x$ & $4.0\times10^{4}$ & 5 & 0.1 & 1.8 & $3\Delta x$ \\
 & $32 $ & $r_{cv}=1.5\Delta x$ & $8.0\times10^{4}$ & 1 & 0.1 & 1.8 & $3\Delta x$ \\
 \tableline
\multirow{4}{*}{OS-RCO-mw} 
& $4 $ & $r_{cv}=1.5\Delta x$ & $1.0\times10^{4}$ & \textbackslash & \textbackslash & \textbackslash & $4\Delta x$ \\
& $8 $ & $r_{cv}=1.5\Delta x$ & $2.0\times10^{4}$ & \textbackslash & \textbackslash & \textbackslash & $3\Delta x$ \\
& $16 $ & $r_{cv}=1.5\Delta x$ & $4.0\times10^{4}$ & \textbackslash & \textbackslash & \textbackslash & $2\Delta x$ \\
& $32 $ & $r_{cv}=1.5\Delta x$ & $8.0\times10^{4}$ & \textbackslash & \textbackslash & \textbackslash & $1\Delta x$ \\
\tableline
\multirow{4}{*}{NS-FNG-mw} 
& $4 $ & $V_{cv}=(16\rm\,pc)^{3}$ & $1.0\times10^{4}$ & 125 & 0.1 & 1.8 & $3\Delta x$ \\
 & $8 $ & $V_{cv}=(16\rm\,pc)^{3}$ & $1.0\times10^{4}$ & 25 & 0.1 & 1.8 & $3\Delta x$ \\
 & $16 $ & $V_{cv}=(16\rm\,pc)^{3}$ & $1.0\times10^{4}$ & 5 & 0.1 & 1.8 & $3\Delta x$ \\
 & $32 $ & $V_{cv}=(32\rm\,pc)^{3}$ & $2.0\times10^{4}$ & 1 & 0.1 & 1.8 & $3\Delta x$ \\
 \tableline
\multirow{4}{*}{NS-RCO-mw} 
 & $4 $ & $V_{cv}=(16\rm\,pc)^{3}$ & $1.0\times10^{4}$ & 125 & 0.1 & 1.8 & $4\Delta x$ \\
 & $8 $ & $V_{cv}=(16\rm\,pc)^{3}$ & $1.0\times10^{4}$ & 25 & 0.1 & 1.8 & $3\Delta x$ \\
& $16 $ & $V_{cv}=(16\rm\,pc)^{3}$ & $1.0\times10^{4}$ & 5 & 0.1 & 1.8 & $2\Delta x$ \\
& $32 $ & $V_{cv}=(32\rm\,pc)^{3}$ & $2.0\times10^{4}$ & 1 & 0.1 & 1.8 & $1\Delta x$ \\
\enddata
\end{deluxetable*}

\section{Results} 
\label{sec:results}
We first analyze the star formation and outflow properties of our highest-resolution ($\Delta x = 4$ pc) simulations in Section 3.1, comparing the NS-FNG-mw, NS-RCO-mw, NS-RCO-gs and OS-FNG-mw models. This comparison focuses on the star formation rate (SFR), cool-phase mass outflow rates, and X-ray luminosities. In Section 3.2, we evaluate the resolution convergence across all seven simulation groups as the grid resolution decreases from $\Delta x = 4$ pc to $32$ pc. 

\subsection{Starburst and outflows properties at high resolution}

\begin{figure*}[!htbp]
    \centering
    \includegraphics[width=0.85\textwidth]{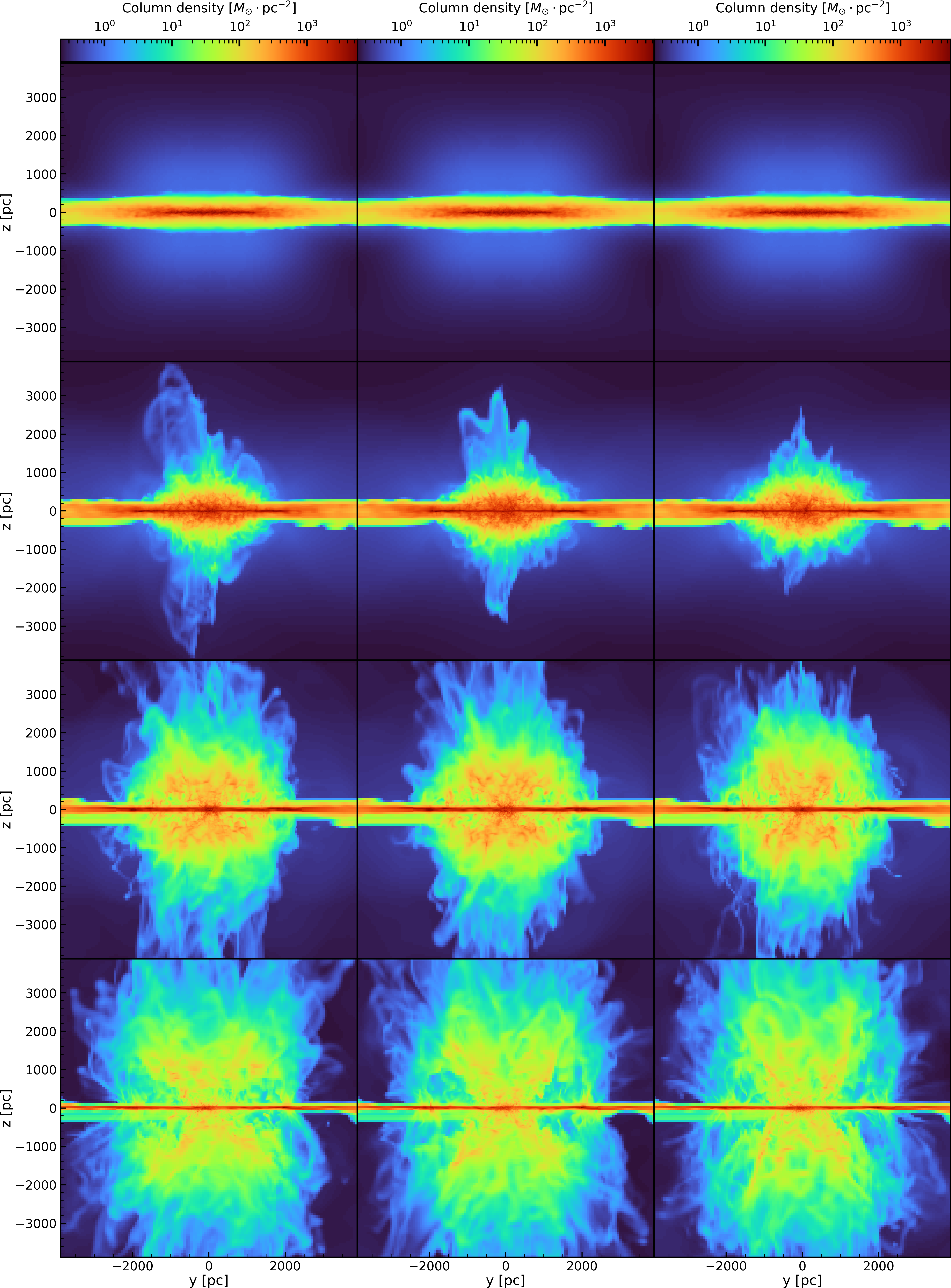}
    \caption{Edge-on projected gas density at $t = 10, 20, 30,$ and $40$ Myr (top to bottom) for the NS-FNG-mw (left), NS-RCO-mw (center), and NS-RCO-gs (right) models at $\Delta x = 4$ pc. } 
    \label{fig:denpro34}
\end{figure*}

Figure~\ref{fig:denpro34} displays the temporal evolution of the edge-on gas column density for the NS-FNG-mw, NS-RCO-mw, and NS-RCO-gs models at $\Delta x = 4$ pc. As shown in the first four rows, galactic outflow develops rapidly between $t = 10$ and $30\rm\,Myr$ across all three runs. The qualitative morphological similarities suggest that, at this resolution, the specific variations in star formation control volumes and SN feedback radii have only a marginal impact on the large-scale structure and development of the outflow.

\begin{figure*}[!htbp]
    \centering
    \includegraphics[width=0.8\textwidth]{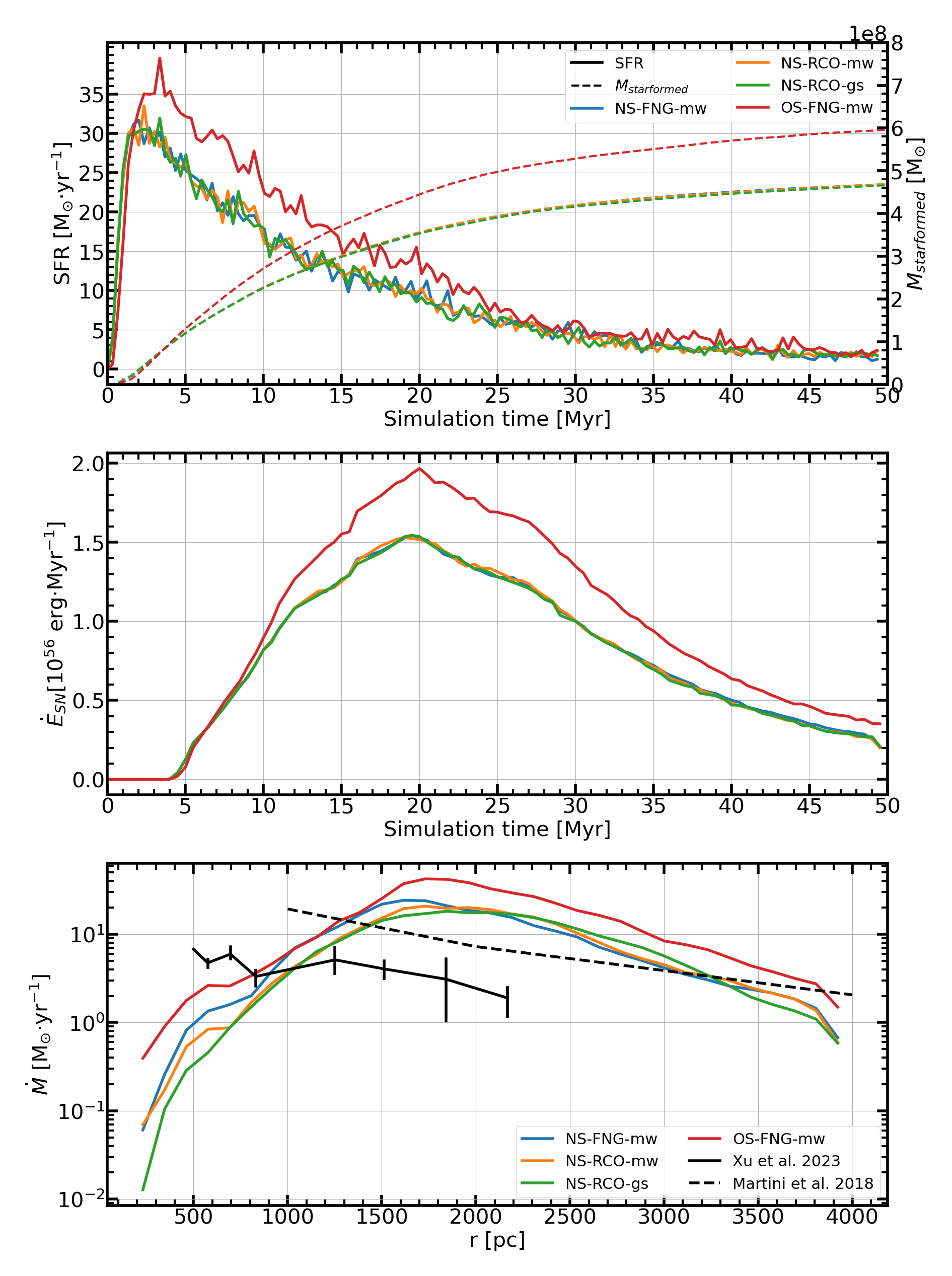}
    \caption{Evolution of star formation and outflow properties. Top: Star formation rate (solid lines) and cumulative stellar mass (dashed lines) as functions of time. Middle: Total energy injection rate from core-collapse supernovae ($\dot{E}_{\rm SN}$) over time. Bottom: Mass outflow rate of the cool gas phase ($\dot{M}_{\rm cool}$) as a function of galactocentric distance at $t = 35\,\text{Myr}$. The different colors represent the four high-resolution models ($\Delta x = 4\,\text{pc}$) as defined in Table~\ref{tab:sim_list}.} 
    \label{fig:sfr_Moutcool_35}
\end{figure*}

Star formation and its associated feedback are the primary drivers of the galactic winds in our simulations. The upper and middle panels of Figure~\ref{fig:sfr_Moutcool_35} show the temporal evolution of the SFR and the corresponding energy injection rate from core-collapse supernovae, $\dot{E}_{\rm SN}$, for the four high-resolution models. Following a short initial latency, SFR rises abruptly, peaking at $\sim 30\,M_{\odot}\,\text{yr}^{-1}$ between $t = 3$ and $5\,\text{Myr}$. This is followed by a sharp decline to approximately $3$--$5\,M_{\odot}\,\text{yr}^{-1}$ by $t \approx 30\,\text{Myr}$, eventually stabilizing at a lower rate of $1.3$--$2.0\,M_{\odot}\,\text{yr}^{-1}$ for $t > 40\,\text{Myr}$. 

The late-stage SFR in our simulations aligns with observational estimates for M82 ($\sim 2.0$--$10.0\,M_{\odot}\,\text{yr}^{-1}$), where the wide range stems largely from IMF uncertainties \citep[e.g.,][]{1998ApJ...498..541K, 2023MNRAS.518.4084Y}. By $t=30\,\text{Myr}$, the NS-FNG-mw, NS-RCO-mw, and NS-RCO-gs groups produce a total stellar mass of $\sim 4.18 \times 10^8\,M_{\odot}$. This is $\sim 25\%$ lower than the reference  OS-FNG-mw model \citep{li2025revisiting}, yet roughly $20\%$ higher than the upper limit of observational estimates for the M82 nuclear starburst ($2.0$--$3.5 \times 10^8\,M_{\odot}$; \citealt{2003ApJ...599..193F, 2009ApJ...697.2030S}). Notably, variations in the SN injection scheme across the three NS groups have a negligible impact on the overall star formation history.

The SN energy injection rate lags the SFR by $\sim5$--$15,\mathrm{Myr}$, consistent with the lifetimes of massive stars before core collapse. It rises more gradually than the SFR, peaks at $\sim20,\mathrm{Myr}$, and then declines slowly. In the groups utilizing the updated star formation and feedback modules, the peak energy injection rate reaches $\sim 1.5 \times 10^{56}\,\text{erg}\,\text{Myr}^{-1}$. This peak is approximately $20\%$ lower than that observed in the reference model from \citet{li2025revisiting}, a result directly attributable to the lower cumulative stellar mass produced by the updated SF prescription.

To enable a consistent comparison of radial outflow properties between simulations and observations, we evaluate the mass outflow rate of the mass-dominant cool phase ($T<2\times10^4\,\mathrm{K}$), $\dot{M}_{\rm cool}(r)$, at $t=35\,\mathrm{Myr}$. We adopt this temperature threshold following \citet{schneider2020physical,schneider2024cgols}. Most stars in the nuclear region of M82 formed within the past $15$--$30\,\mathrm{Myr}$, with a current SFR of $\sim2$--$3\,M_\odot,\mathrm{yr}^{-1}$ \citep{2003ApJ...599..193F}. At $t=35\,\mathrm{Myr}$, our simulations are $\sim30\,\mathrm{Myr}$ past the SFR peak, with the SFR declining to $\sim2$--$4\,M_\odot,\mathrm{yr}^{-1}$ and the SN energy injection rate to $\sim1/3$ of its peak value. Meanwhile, the galactic wind has developed to kiloparsec scales, following the rapid growth of the outflow at $t\sim10$--$30\,\mathrm{Myr}$ (Figure~\ref{fig:denpro34}). We therefore adopt $t=35\,\mathrm{Myr}$ as a representative epoch, corresponding to a mature starburst phase broadly comparable to M82.

We emphasize, however, that the outflow is still evolving at $t=35\,\mathrm{Myr}$ despite the declining SN rate. This choice should therefore be regarded as a representative comparison epoch rather than an indication of a steady state. To illustrate the time dependence, we also present the same radial cool-phase outflow profiles at several nearby epochs in Appendix~\ref{sec:Mout_othertime}, which show that the outflow properties continue to evolve at $t=35\,\mathrm{Myr}$.

The bottom panel of Figure~\ref{fig:sfr_Moutcool_35} presents $\dot{M}_{\rm cool}$ measured at $t = 35\,\text{Myr}$ as a function of galactocentric radius, following the methodology in Section 4.2 of \citet{li2025revisiting} with a radial velocity criterion of $v_{\rm rad} > 0\,\text{km\,s}^{-1}$. In our simulations utilizing the upgraded SF module, $\dot{M}_{\rm cool}$ increases gradually to a peak of $\sim 20\,M_{\odot}\,\text{yr}^{-1}$ at $r \sim 1.6\,\text{kpc}$ before declining to below $1\,M_{\odot}\,\text{yr}^{-1}$ beyond $r = 3.5\,\text{kpc}$. This radial profile is broadly consistent with the "OS" model from \citet{li2025revisiting}, though the upgraded simulations exhibit moderately lower $\dot{M}_{\rm cool}$ at $r < 1\,\text{kpc}$ and $r > 1.5\,\text{kpc}$.

Observational estimates of $\dot{M}_{\rm cool}$ from \citet[ionized gas; $T\sim 10^4\,\text{K}$]{xu2023radial} and \citet[neutral atomic hydrogen]{2018ApJ...856...61M} are included in the bottom panel for comparison. We note that \citet{2018ApJ...856...61M} measured the outflow rate as a function of projected distance along the minor axis; the corresponding three-dimensional radial profile is expected to be flatter. Furthermore, their rates may be overestimated, as discussed in \citet{2023MNRAS.518.4084Y}. At a resolution of $4\,\text{pc}$, our simulated $\dot{M}_{\rm cool}$ is broadly consistent with the combined results from these studies. For $r \gtrsim 1.0\,\text{kpc}$, the simulated radial trend shows excellent agreement with observations. In the inner region ($r \lesssim 1.0\,\text{kpc}$), $\dot{M}_{\rm cool}$ in most models remains below the estimates from \citet{xu2023radial}, a discrepancy that may be partially attributed to projection effects in the observational data.


We further evaluate the properties of the hot outflow phase by generating synthetic broadband ($0.5$--$7.0\,\text{keV}$) X-ray emission maps using the \texttt{pyXSIM} \footnote{https://hea-www.cfa.harvard.edu/{\textasciitilde}jzuhone/pyxsim/index.html} package. Thermal emission is modeled using the APEC collisional ionization equilibrium (CIE) framework, with the metallicity fixed to the volume-averaged value of gas exceeding $10^{6}\,\text{K}$ within the simulation domain. \citet{lopez2020temperature} reported a total X-ray luminosity of $3.90 \times 10^{40}\,\text{erg}\,\text{s}^{-1}$ for M82 in the $0.5$--$7.0\,\text{keV}$ band based on \textit{Chandra} observations. Accounting for contributions from point sources, charge exchange, and non-thermal components (estimated at $\sim 30\%$), the thermal X-ray luminosity specifically attributable to hot diffuse gas is $\sim 2.73 \times 10^{40}\,\text{erg}\,\text{s}^{-1}$. To ensure a consistent comparison, we extract the mock X-ray luminosity from our simulations within the same spatial footprint used in the observational analysis (for a detailed description of the aperture, see \citealt{li2025revisiting}).

\begin{table}[htbp]
        \tablenum{3}
	\centering
	\caption{Thermal X-ray luminosities ($L_{X, \rm{tot}}$) at $t = 35$ Myr and their ratios to the M82 observational value for four  simulations with $\Delta x = 4$ pc.}
	\begin{tabular}{lccr}
		\hline
		Simulation & $L_{X,tot}$ (erg/s) & Ratio to observation \\
		\hline
		NS-FNG-mw& $2.178\times 10^{40} $ &0.836  \\
        NS-RCO-mw& $2.006\times 10^{40} $ & 0.853  \\
        NS-RCO-gs& $1.649\times 10^{40} $ & 0.764  \\
        OS-FNG-mw& $3.436\times 10^{40} $ & 1.258  \\
		\hline
	\end{tabular}
	\label{tab:Lx} 
\end{table}

\begin{figure}[htbp]
    \centering
    \includegraphics[width=0.95\columnwidth]{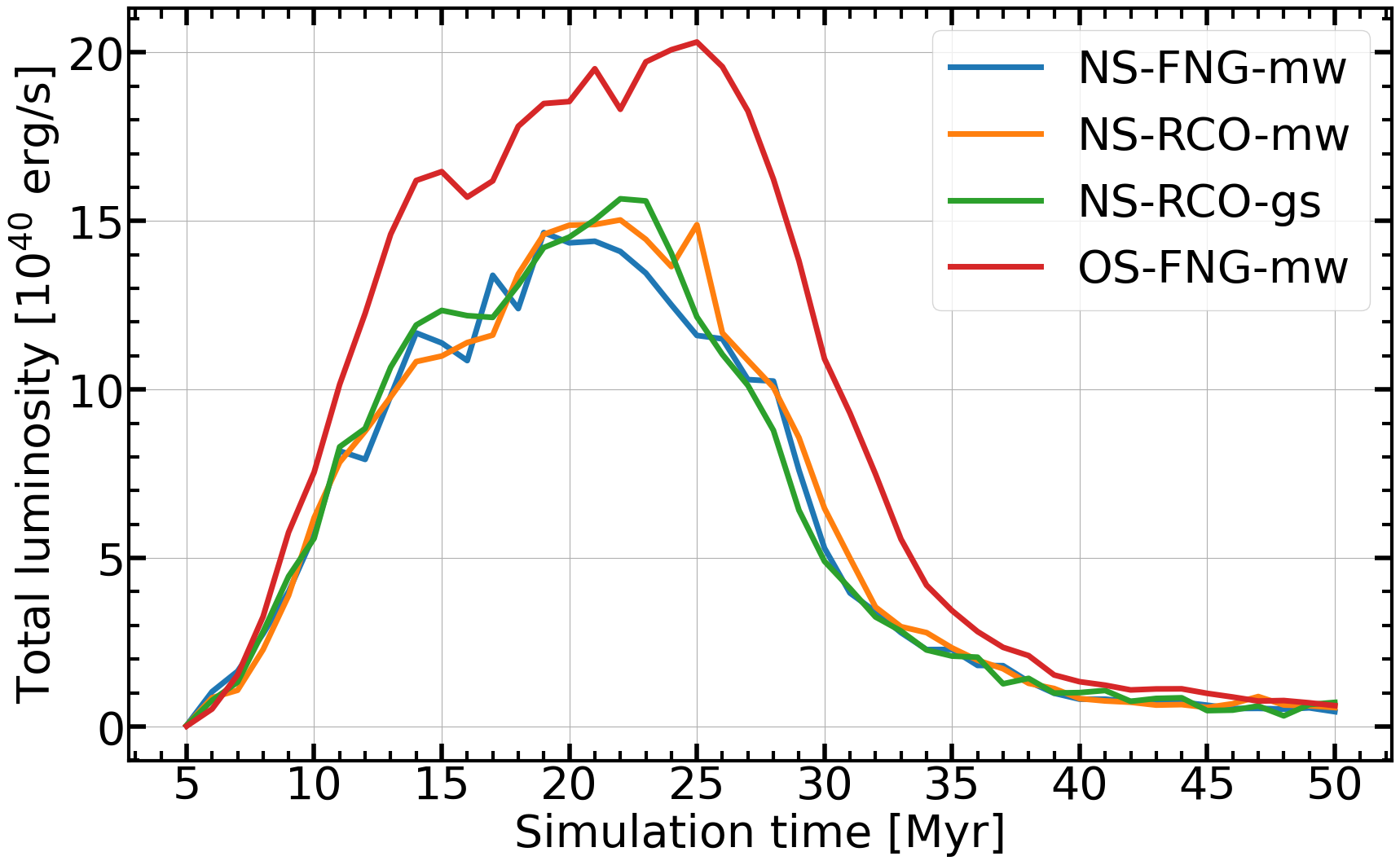}
    \caption{Temporal evolution of the total thermal X-ray luminosity, $L_{X, \rm{tot}}$, for the hot gas phase across four high-resolution simulations.}
    \label{fig:xray_t}
\end{figure}

We compare $L_{\rm X,tot}$ at $t=35\,\mathrm{Myr}$ with observations, consistent with the epoch adopted by cool-phase outflow comparison above. The total X-ray luminosities, $L_{X,\rm tot}$, of the hot outflow phase at $t = 35\,\text{Myr}$ are summarized in Table~\ref{tab:Lx}, showing good agreement with observational data. Figure~\ref{fig:xray_t} illustrates the temporal evolution of $L_{X,\rm tot}$ for the simulations at $\Delta x = 4\,\text{pc}$. The luminosity increases from negligible levels at $t = 5\,\text{Myr}$ to a peak between $t = 20$ and $25\,\text{Myr}$. The hot gas maintains a substantial X-ray luminosity throughout the epoch of $10$--$35\,\text{Myr}$, coinciding with the rapid development of the multiphase outflow as described in \citet{li2025revisiting}. Expanding the SN feedback deposition radius from $3\Delta x = 12\,\text{pc}$ (the FNG'' scheme) to $\sim 20\,\text{pc}$ (the RCO'' scheme) results in a modest increase in peak luminosity during the $20$--$25\,\text{Myr}$ interval; however, it does not significantly alter the long-term evolutionary trend for $t \gtrsim 30\,\text{Myr}$ ( the blue curve in Figure~\ref{fig:xray_t}).

In summary, at a resolution of $4\,\mathrm{pc}$, our simulations broadly reproduce the key observed starburst characteristics and multiphase outflow properties of M82, while the updated star formation scheme yields a cumulative stellar mass in better agreement with observations.

\subsection{Resolution Convergence}

In this section, we assess the resolution convergence of our star formation and feedback prescriptions using simulations with $\Delta x=4$, $8$, $16$, and $32\,\mathrm{pc}$. We first examine four integrated properties: the cumulative stellar mass, $M_\star$; total kinetic energy; peak mass outflow rates of the cool, warm, and hot phases, $\dot{M}_{\rm max}$; and total thermal X-ray luminosity, $L_{X,\rm tot}$. We then examine normalized wind properties, i.e., the mass and energy loading factors and X-ray emission efficiency, to distinguish variations in the overall star formation and SN feedback budgets from those in wind-driving efficiency.

To quantitatively assess convergence, we define the relative error of a quantity compared with the highest-resolution (4 pc) simulation as
\begin{equation}
    {\rm RE4} =
    \frac{|a-a_{4\,{\rm pc}}|}
         {a_{4\,{\rm pc}}}\times100\%,
\end{equation}
where $a_{4\,{\rm pc}}$ is the value measured in the 4 pc simulation. Unless otherwise specified, relative errors at 8, 16, and 32 pc are evaluated relative to the 4 pc run within the same model configuration.

\subsubsection{Stellar Mass Produced in the Starburst}

\begin{figure*}[!htbp]
    \centering
    \includegraphics[width=1\textwidth]{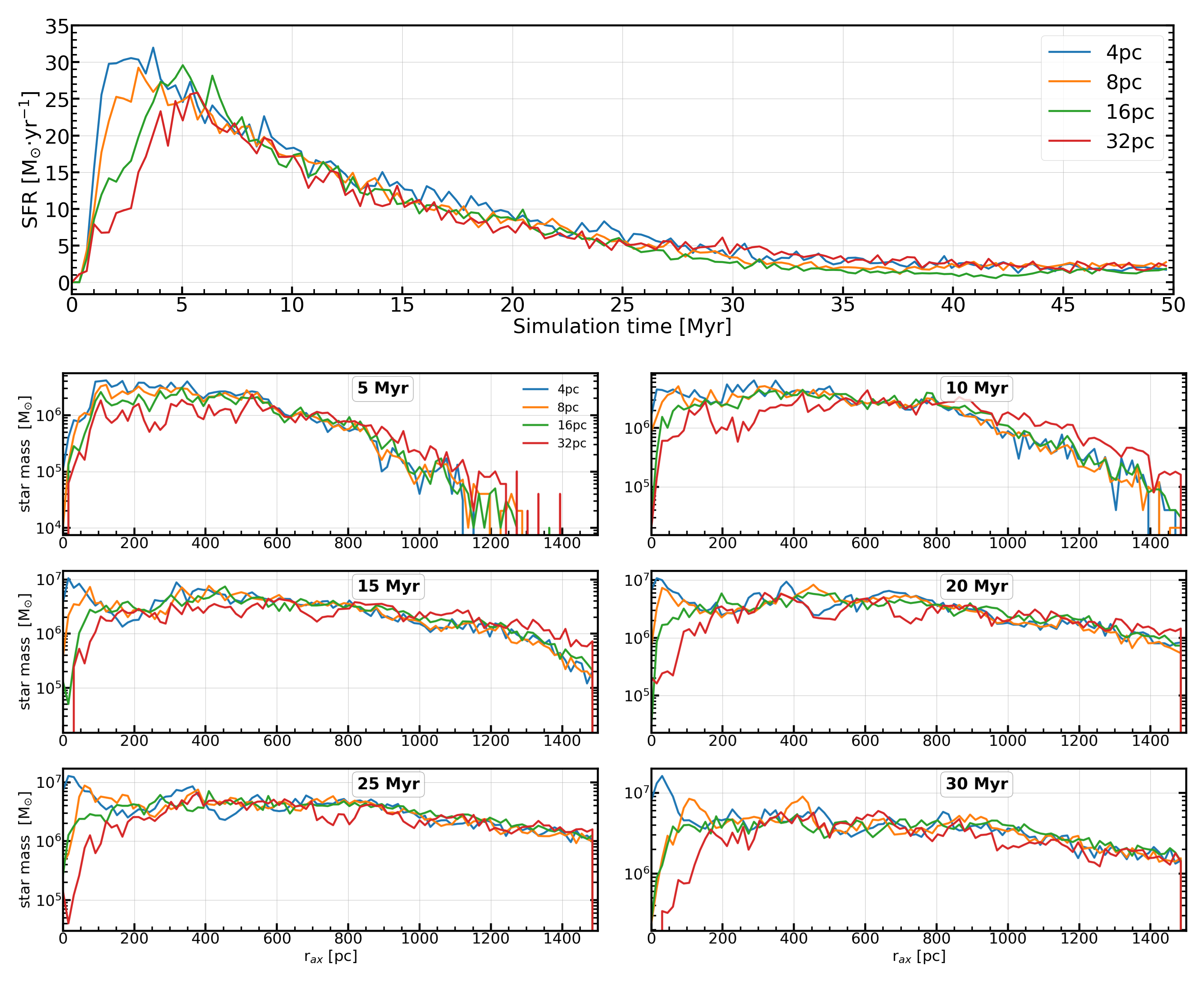}
    \caption{The star formation rate of NS-RCO-gs and the corresponding formed star mass distribution along the cylindrical radius $r_{\rm{ax}}=\sqrt{x^{2}+y^{2}}$ at different time.}
    \label{fig:mw3sfdis}
\end{figure*}

\begin{figure*}[!htbp]
    \centering
    \includegraphics[width=1\textwidth]{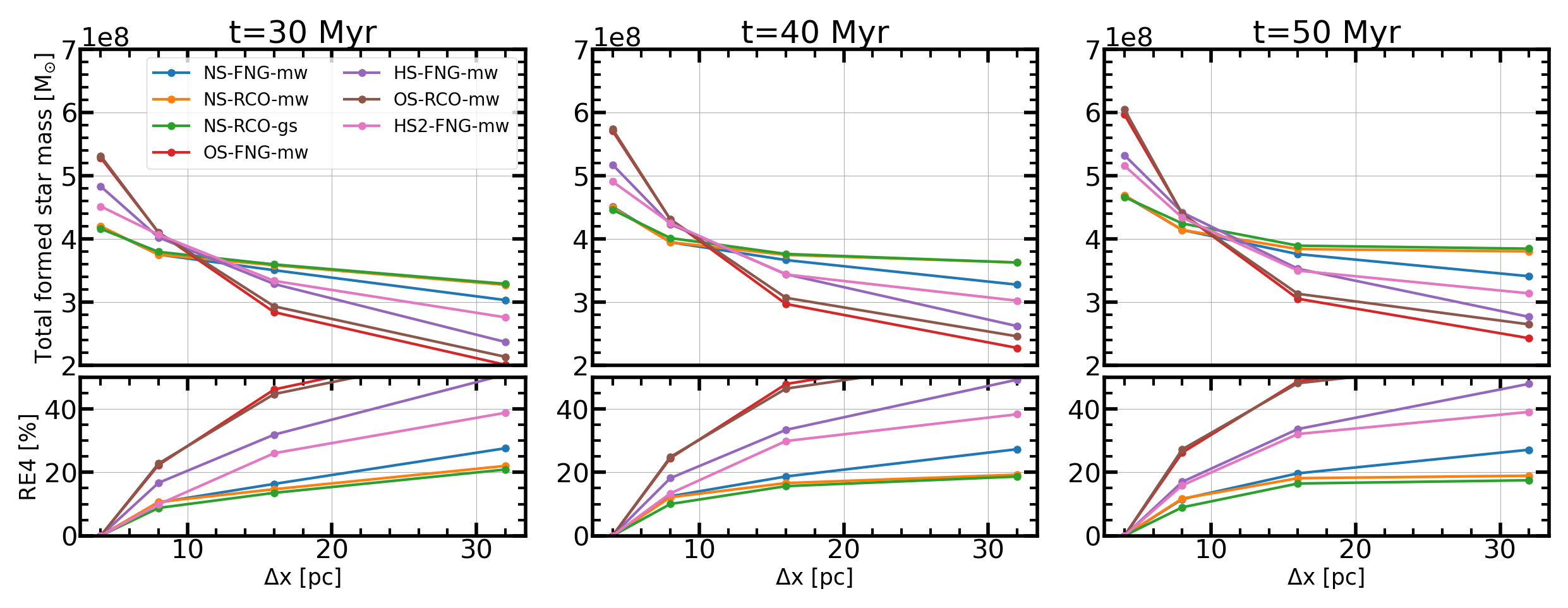}
    \caption{Cumulative stellar mass as a function of resolution at $t = 30, 40,$ and $50\rm\,Myr$. Top: Absolute values across different sub-grid models. Bottom: Relative error (${\rm RE}4$) for resolutions from 8 to 32,pc. ``OS'' denotes models with a control volume of $1.5\Delta x$ for accretion; ``HS'' denotes models with fixed control volume accretion; "NS" denotes the updated recipes using fixed control volume accretion with resolution-dependent efficiency. ``FNG'' and ``RCO'' indicate feedback deposition radii fixed to a multiple of the grid size ($3 \Delta x$) or scaled to the physical cooling radius ($R_{\rm cool}$), respectively. The suffixes ``mw'' and ``gs'' distinguish between mass-weighted and Gaussian-weighted injection kernels.} 
    \label{fig:sf_diff}
\end{figure*}

We first provide an overview of star formation across the tested resolution range. The upper panel of Figure~\ref{fig:mw3sfdis} presents the star formation history (SFH) for the NS-RCO-gs group, which is highly consistent across resolutions. However, coarser resolutions exhibit a slightly delayed peak in the star formation rate, primarily because lower-resolution simulations require more time to develop the high-density gas peaks necessary for star formation. Following $t \gtrsim 5\text{ Myr}$, the SFH achieves approximate convergence across all resolutions.

The bottom panel of Figure~\ref{fig:mw3sfdis} illustrates the radial distribution of stellar mass formed at intervals between 5 and 30 Myr. Here, the x-axis represents the cylindrical radius $r_{\rm ax} = \sqrt{x^{2}+y^{2}}$ from the disk center (up to $1500\text{ pc}$), and the y-axis shows the stellar mass within bins of $\Delta r = 15\text{ pc}$. The overall spatial distribution is highly consistent; however, minor discrepancies arise within the central regions of the disk, where stellar mass tends to be more concentrated at higher resolutions.

The upper panel of Figure~\ref{fig:sf_diff} illustrates the cumulative stellar mass produced as a function of spatial resolution at $t = 30, 40,$ and $50\,\text{Myr}$. Each line represents a simulation suite sharing identical star formation and SN feedback sub-grid configurations. The lower panel quantifies the change in cumulative stellar mass with coarsening resolution using the relative error RE4.

The original ``OS'' recipe \citep{Wang2026, li2025revisiting} produces a rapid decline in cumulative stellar mass at coarser resolutions, with RE4 reaching $\sim60\%$ at $\Delta x=32$ pc. In contrast, the cumulative stellar mass in ``NS'' models remain remarkably stable from $4$ to $32$ pc, with ``NS-RCO'' maintaining RE4 below $\sim20\%$ across the full resolution range. Individually, adopting a nearly fixed control volume flattens the decline, reducing the RE4 from $\sim60\%$ in ``OS'' to $\sim50\%$ in ``HS'' at $\Delta x=32$ pc. Using a resolution- and density-dependent star formation efficiency reduces it to $\sim40\%$ in ``HS2''. Changing the feedback coupling region from $3\Delta x$ to $\sim R_{\rm cool}$ provides a modest additional improvement, while replacing ``mw'' with ``gs'' weighting yields a small further reduction. 

\subsubsection{Kinetic Energy budget}

\begin{figure*}[!htbp]
    \centering
    \includegraphics[width=1\textwidth]{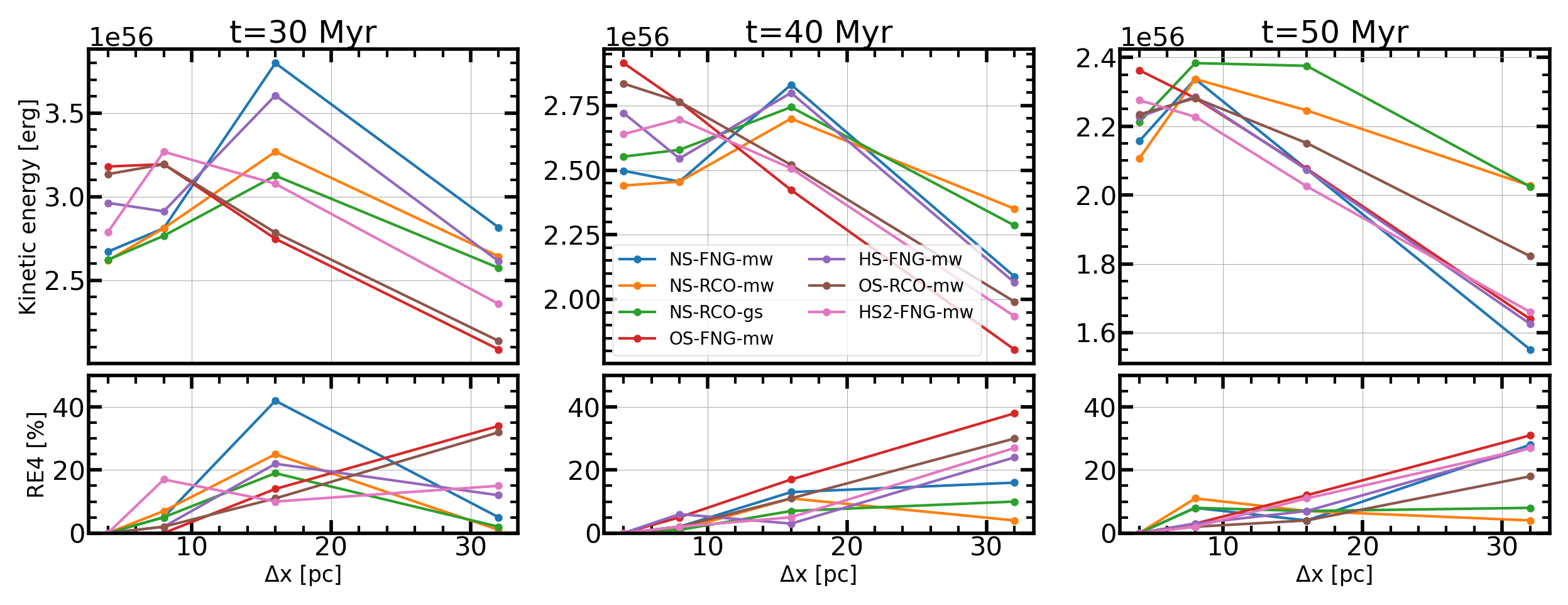}
    \caption{Total kinetic energy versus resolution at $t = 30, 40,$ and $50\rm\,Myr$. Top: Integrated kinetic energy within the simulation domain. Bottom: Relative error (${\rm RE}4$) for resolutions up to $32$,pc. Different colors and markers share the same meaning as Figure \ref{fig:sf_diff}.}
    \label{fig:kne_diff}
\end{figure*}

Core-collapse supernova feedback is the primary engine driving galactic outflows in low-mass galaxies. Physically, the energy injection typically  experience three stages: the initial free expansion of ejecta, the adiabatic (Sedov-Taylor) phase—characterized by conserved energy and negligible radiative losses—and the subsequent radiative phase, where cooling dominates and momentum is conserved. Thereby, a significant fraction of the injected energy is converted into the kinetic energy of the swept-up and shock-accelerated gas. This kinetic energy reservoir is a critical determinant of outflow properties, the turbulent state of the interstellar medium (ISM), and subsequent star formation activity.

The upper panels of Figure~\ref{fig:kne_diff} show the total kinetic energy within the simulation domain at $t=30$, $40$, and $50$ Myr for different models and resolutions, with the corresponding RE4 errors shown in the lower panels. The reference OS-FNG-mw model, which adopts the feedback scheme of \citet{kim2017three} and used in \citet{li2025revisiting}, also shows a rapid decline in kinetic energy as the resolution coarsens, with RE4 exceeding $25\%$ at $\Delta x=32$ pc. These errors are  smaller than those in the cumulative stellar mass, which determines the total SN feedback energy. This difference arises because the \citet{kim2017three} scheme increases the fraction of momentum injection at coarser resolutions (see Section~5.2), partially compensating for the reduced stellar mass and associated CCSN energy input.

The total kinetic energy in our fiducial NS-RCO-gs model fluctuates moderately with resolution, but its RE4 remains below $\sim20\%$ across $4$--$32$ pc and is generally smaller than the corresponding stellar-mass error. Other models show a similar trend, reflecting the compensating effect of increased momentum injection at coarser resolution. However, when the stellar-mass decline is modest, a rapid increase in the momentum fraction can temporarily boost the kinetic energy, producing larger kinetic-energy than stellar-mass errors in NS-FNG-mw, NS-RCO-mw, and NS-RCO-gs at $\Delta x=16$ pc when $t=30$ Myr. This effect is strongest in NS-FNG-mw because FNG produces a larger variation in the kinetic-injection fraction across different resolutions than RCO (Section~5.2).

\subsubsection{Mass outflow rate}
\label{sec:Mass outflow rate of cool phase}
\begin{figure*}[!htbp]
    \centering
    \includegraphics[width=1\textwidth]{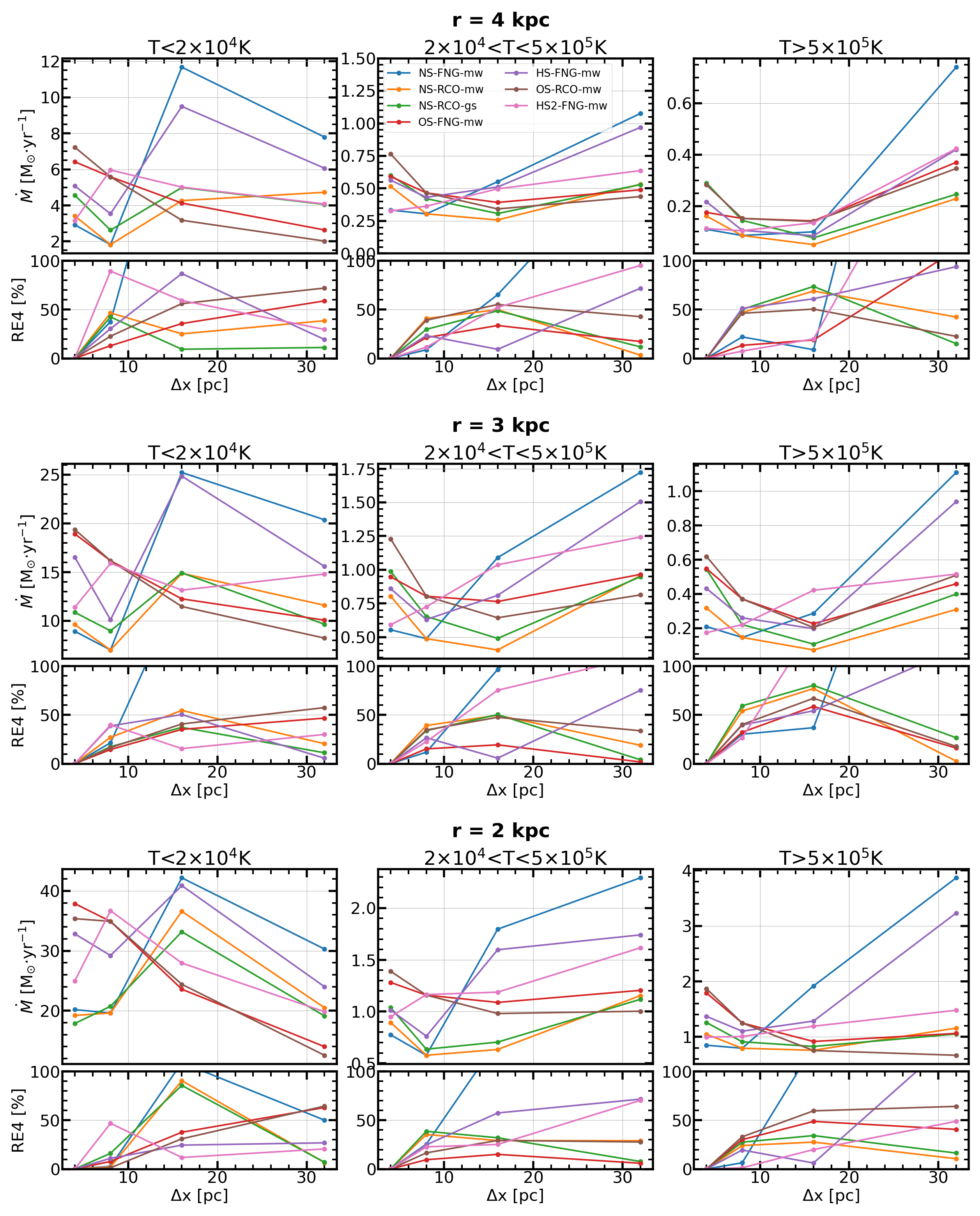}
    \caption{Maximum mass outflow rates for the three gas phases as a function of simulation resolution, measured at $r=4,3,2\text{ kpc}$. The upper panel shows the absolute outflow rates, while the lower panel illustrates the relative errors (normalized to the $4\text{ pc}$ run) across the 4–32 pc resolution range for each simulation group.}
    \label{fig:Mout_diff}
\end{figure*}

The impact of galactic winds on host galaxy evolution, particularly in regulating long term star formation, is primarily determined by the mass outflow rate. A defining characteristic of these winds is their multiphase structure. The interaction between cold and hot phases, driven by processes such as turbulent mixing and radiative cooling, fundamentally shapes the wind's global properties \citep{2022ApJ...924...82F}. In galaxy-scale hydrodynamic simulations, limited resolution often leads to numerical smoothing, which hinders the accurate capture of these small-scale interactions. This resolution dependence can result in the loss of multiphase detail, ultimately biasing the predicted outflow rates and their subsequent feedback on the galaxy.

We therefore evaluate resolution convergence for the cool, warm, and hot phase mass outflow rates across each sub-grid configuration. Direct comparisons at fixed radii or times are often complicated by varying wind development timescales \citep{li2025revisiting}. To isolate the intrinsic driving capacity of the SN feedback, we focus on the maximum mass outflow rate ($\dot{M}_{\rm max}$) at specific galactocentric distances, independent of the exact time it occurs.


We first examine the mass outflow rates measured at a galactocentric radius of $r = 4\,\mathrm{kpc}$. This choice is motivated by two considerations. First, outflow properties at this radius are less affected by the central starburst region, where strong spatial and temporal fluctuations arise from the coupled effects of star formation and feedback injection, particularly at $z \lesssim 1.2\,\mathrm{kpc}$ \citep{li2025revisiting}. Second, gas at $r = 4\,\mathrm{kpc}$ is largely decoupled from the galactic disk, allowing it to more directly probe the role of outflows in redistributing the ISM and regulating star formation over longer timescales. For completeness, we also present the corresponding results at $r = 3$ and $2\,\mathrm{kpc}$.

For the reference OS-FNG-mw model, the cool phase, which dominates the total outflow rate, declines systematically with coarsening resolution and shows poor convergence. Its RE4 increases from $\sim40\%$ at $\Delta x=16$ pc to $\sim60\%$ at $\Delta x=32$ pc, comparable to the cumulative stellar-mass errors. The warm phase is better converged, with RE4 below $\sim30\%$ at $\Delta x\geq16$ pc, whereas the RE4 error of hot phase increase gradually with $\Delta x$ to above $100\%$ at $\Delta x=32$ pc.

In contrast, $\dot{M}$ in our fiducial NS-RCO-gs model fluctuates moderately with resolution, yielding substantially improved convergence for the cool and total phases. For the cool phase, RE4 is $\lesssim40\%$ at $\Delta x=8$ pc and $\lesssim10\%$ at $\Delta x\geq16$ pc, well below the corresponding cumulative stellar-mass errors. The warm phase is slightly less well converged than in OS-FNG-mw, but remains at RE4 $\lesssim50\%$ across 4--32 pc, while the hot-phase RE4 remains $\lesssim70\%$.

The HS-FNG-mw and HS2-FNG-mw models, which modify only the star-formation prescription while retaining the original feedback coupling, improve convergence at some resolutions but do not reduce the overall variation. Changing the injection radius to $R_{\rm cool}$ alone also fails to improve the convergence of the peak mass outflow rate ($\dot{M}_{\rm max}$) in OS-RCO-mw; OS-FNG-mw actually shows better convergence for the cool and warm phases. Moreover, combining the modified star-formation prescription with the original feedback scheme in NS-FNG-mw results in poorer convergence, with RE4 substantially exceeding the corresponding cumulative stellar-mass error.

Within the NS model family, tying the SN injection radius to $R_{\rm cool}$ (NS-RCO-mw and NS-RCO-gs) substantially improves convergence relative to NS-FNG-mw, reducing the RE4 of the cool-phase and total mass outflow rates at $r=4\,\rm kpc$ from $\gtrsim100\%$ to $\lesssim40\%$. The Gaussian-weighted scheme (NS-RCO-gs) provides a modest further improvement over the mass-weighted scheme (NS-RCO-mw) for the cool phase, while the two schemes show comparable convergence for the warm and hot phases.

We further evaluate the peak mass outflow rates at $r=3$ and $2\,\mathrm{kpc}$, with the results shown in the middle and bottom rows of Figure~\ref{fig:Mout_diff}. Overall, the convergence at $r=2$ and $3\,\mathrm{kpc}$ is comparable to, or slightly poorer than, that at $r=4\,\mathrm{kpc}$ for most simulation groups. The NS-RCO-mw and NS-RCO-gs runs maintain good agreement across resolutions, with the main exception being the cool-phase outflow rate at $r=2\,\mathrm{kpc}$ in the $\Delta x=16\,\mathrm{pc}$ simulations. This discrepancy may arise from a combination of resolution-dependent cooling and enhanced phase mixing near this radius. While the origin of this feature remains uncertain, it highlights limitations of the current model that warrant further investigation and refinement in future work.

\begin{figure*}[!htbp]
    \centering
    \includegraphics[width=0.85\textwidth]{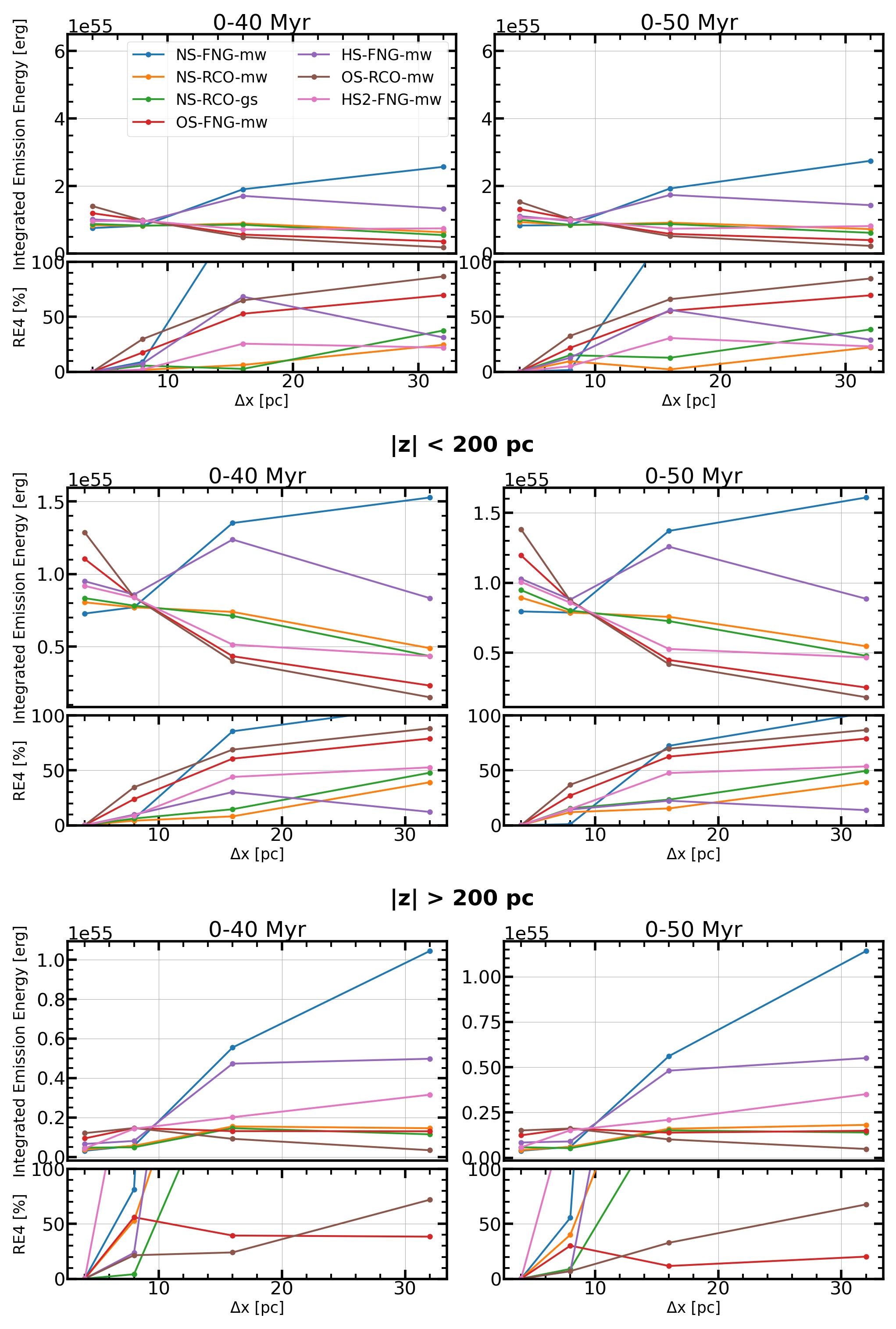}
    \caption{Time-integrated X-ray emission over $t=0$--$40$ Myr (left column) and $0$--$50$ Myr (right column) as a function of simulation resolution, measured over the entire simulation domain (top row), the disk region ($|z|<200\,\rm pc$; middle row), and the outflow region ($|z|\ge200\,\rm pc$; bottom row). The upper panel in each row displays absolute values, while the lower panel shows the relative errors with respect to resolution of 4 pc within the same simulation group.}
    \label{fig:xray_error}
\end{figure*}

\subsubsection{X-ray Emission}

Hot-phase gas ($T > 10^6\rm\,K$) is the primary source of X-ray emission in starburst galaxies and a hallmark of starburst-driven outflows. Since X-ray luminosity ($L_X$) results from the thermalization of supernova energy, reproducing observed emission levels is a critical benchmark for feedback models. Furthermore, because X-ray emissivity in collisional ionization equilibrium (CIE) scales as $L_X \propto \int n^2 \Lambda(T, Z) dV$, its convergence across resolutions serves as a sensitive indicator of how well the simulation resolves the density and temperature structure of the multiphase gas.


The top row of Figure~\ref{fig:xray_error} shows the integrated X-ray emission over $t=0$--$40$ Myr and $0$--$50$ Myr across the simulation domain, together with the corresponding relative errors (RE4). The $0$--$40$ Myr interval covers the main X-ray emission phase (Figure~\ref{fig:xray_t}), whereas $0$--$50$ Myr covers the full simulation period, including the low X-ray luminosity phase after $t\sim40\,\rm Myr$. Comparing the two intervals tests whether the resolution convergence is sensitive to this late-time phase. For the reference OS-FNG-mw model, the integrated X-ray emission decreases with coarsening resolution, yielding RE4 $\sim70\%$ at $\Delta x=32$ pc for both time intervals, larger than the corresponding stellar-mass error. In contrast, the fiducial NS-RCO-gs model exhibits substantially improved resolution convergence, with RE4 below $\sim40\%$ across $4$--$32$ pc for both intervals, although still larger than the corresponding stellar-mass error.

Both HS-FNG-mw and HS2-FNG-mw improve convergence relative to OS-FNG-mw, with HS2-FNG-mw showing the stronger improvement, highlighting the importance of the revised SFE scheme. In contrast, OS-RCO-mw performs worse than OS-FNG-mw, indicating that modifying the feedback coupling radius alone does not improve X-ray emission convergence.

To further investigate whether the convergence of X-ray emission is influenced by the SMR structure outside the disk, we decompose the cumulative X-ray emission into a central disk component ($|z|<200\,\mathrm{pc}$) and an outer outflow component ($|z|>200\,\mathrm{pc}$), as shown in the middle and bottom rows of Figure~\ref{fig:xray_error}. Observations indicate that the disk region dominates the total X-ray emission \citep{lopez2020temperature}. For the NS-RCO-mw and NS-RCO-gs models, the convergence of the disk emission is comparable to, or slightly worse than, that of the total emission, with relative errors remaining below $\sim40\%$ and $\sim50\%$, respectively, over the $4$--$32\,\mathrm{pc}$ resolution range, for both time intervals.

In contrast, the X-ray emission from the outflow region shows substantially larger divergence, except in the OS-FNG-mw runs. In most models, the X-ray emission above disk remains relatively weak when the resolution in the starburst region is finer than $10\,\mathrm{pc}$, but increases sharply once it becomes coarser than this resolution. This behavior likely reflects a combination of effects. The density of hot outflowing gas decreases rapidly with galactocentric distance, while higher resolution better resolves the multiphase structure of the hot phase. As a result, the hot gas content at $z \ge 200\,\mathrm{pc}$ is more sensitive to resolution changes than in the central disk region, particularly given the progressively coarser resolution with height in our SMR setup.

\subsubsection{Loading factors and X-ray emission efficiency}

The absolute outflow properties depend on both the total stellar mass formed and the associated SN feedback energy budget, as well as the efficiency of converting feedback energy into wind motion. Thus, variations in the outflow rate and X-ray luminosity may reflect changes in either the overall feedback budget or the wind-driving efficiency. In principle, normalizing these quantities by the SFR or SN feedback energy budget can remove their dependence on the overall feedback budget and provide a more direct measure of resolution convergence in wind-driving efficiency. We therefore consider the commonly used mass-loading factor, energy-loading factor, and X-ray emission efficiency, given by
\begin{equation}
\eta_M(t)=\frac{\dot{M}_{\rm out}(t)}{\mathrm{SFR(t)}},
\end{equation}
\begin{equation}
\eta_E(t)=\frac{\dot{E}{\rm out}(t)}{\dot{E}_{\rm SN}(t)}, 
\end{equation}
and
\begin{equation}
\epsilon_{L_X}(t)=\frac{L_X(t)}{\dot{E}_{\rm SN}(t)},
\end{equation}
respectively. For $\eta_M$ and $\eta_E$, we consider the total outflow, including the cool, warm, and hot phases, because substantial mass and energy exchange occurs between these phases. The total outflow therefore provides a more representative measure of the overall wind-driving efficiency. We evaluate these two quantities at $r=4\,\rm kpc$, following the motivation given in Section~3.2.3. For $L_X(t)$, we consider the total X-ray luminosity of hot gas in the disk and outflow regions. 

Moreover, \citet{li2025revisiting} and \citet{2026ApJ..1000..158C} showed that outflow properties can vary strongly with time and spatial scales, with time lags between star formation, SN energy injection, and the resulting outflow properties. These lags depend on both the measured property and height above the gas disk. Moreover, Figure ~\ref{fig:sfr_Moutcool_35} and ~\ref{fig:xray_t} show distinct temporal profiles for the SFR, SN energy injection rate, and X-ray luminosity, including different rise and decline timescales. Together with the bursty nature of the SFR, these effects make instantaneous normalized wind properties unreliable measures of wind-driving efficiency. To mitigate them, we average $\eta_M$ and $\eta_E$ at $r=4\,\rm kpc$ over $30$--$50\,\mathrm{Myr}$, when the large-scale outflow has fully developed there, and $\epsilon_{L_X}$ over $10$--$30\,\mathrm{Myr}$, when the X-ray emission is well established, as follows:

\begin{equation}
\bar\eta_M=
\frac{\langle\dot{M}_{\rm out}(t)\rangle_{30-50\,\rm Myr}}
{\langle\mathrm{SFR}(t)\rangle_{5-25\,\rm Myr}},
\label{eqn:ave_mass_load}
\end{equation}

\begin{equation}
\bar\eta_E=
\frac{\langle\dot{E}_{\rm out}(t)\rangle_{30-50\,\rm Myr}}
{\langle\dot{E}_{\rm SN}(t)\rangle_{20-40\,\rm Myr}},
\label{eqn:ave_ene_load}
\end{equation}

and

\begin{equation}
\bar\epsilon_{L_X}=
\frac{\langle L_X(t)\rangle_{10-30\,\rm Myr}}
{\langle\dot{E}_{\rm SN}(t)\rangle_{5-25\,\rm Myr}}
\label{eqn:ave_x_eff}
\end{equation}

Note that, in Equations~\ref{eqn:ave_mass_load} , ~\ref{eqn:ave_ene_load}, we introduce time offsets when averaging the SFR and SN energy injection rate to account for the corresponding time lags at $r=4\,\rm kpc$. In Equation~\ref{eqn:ave_x_eff}, we similarly introduce a time offset to account for the lag between the peak of SN energy injection and the resulting X-ray emission (Figure~\ref{fig:sfr_Moutcool_35} and ~\ref{fig:xray_t}). We calculate these quantities for each resolution and simulation group and evaluate the corresponding relative error, RE4. The results are presented below.

\paragraph{Mass loading factor}

\begin{figure*}[!htbp]
    \centering
    \includegraphics[width=1\textwidth]{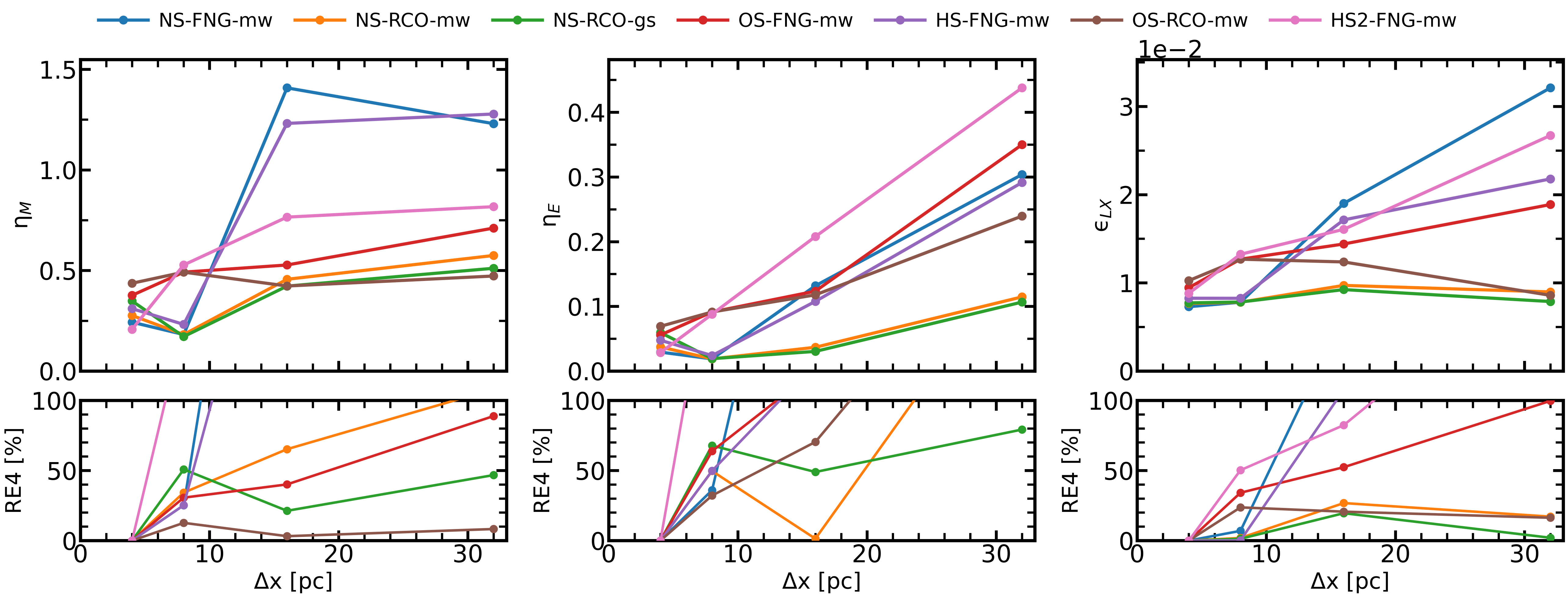} 
    \caption{The time-averaged mass (left), energy (middle) loading factors at 4 kpc height and, the time-averaged X-ray efficiency (right) in the entire simulation area. The upper panel in each row displays absolute values, while the lower panel (with zoom-in) shows the relative errors with respect to resolution of 4 pc within the same simulation group.}
    \label{fig:eta_all}
\end{figure*}

The left panel of Figure~\ref{fig:eta_all} shows $\bar\eta_M$ averaged over $30$--$50\,\mathrm{Myr}$ and the corresponding relative errors at $r=4\,\rm kpc$. For the reference OS-FNG-mw model, $\bar\eta_M$ increases from $\sim 0.35$ to $\sim 0.7$ as the resolution coarsens, likely due to the increasing fraction of momentum injection (Section~5.2). Consequently, RE4 increases from $\sim30\%$ at $\Delta x=8$ pc to $\sim40\%$ at $16$ pc and $\sim90\%$ at $32$ pc. In contrast, $\bar\eta_M$ in NS-RCO-gs shows much weaker resolution dependence, fluctuating around $\sim 0.4$, with RE4 ranging from $\sim25\%-50\%$.

Modifying only the sink control volume (HS-FNG-mw) or star formation efficiency (HS2-FNG-mw) does not improve the convergence of $\bar\eta_M$ relative to OS-FNG-mw. Similarly, NS-FNG-mw shows poor convergence despite the whole modified star formation scheme. In contrast, adopting an SN coupling radius close to $R_{\rm cool}$ substantially improves convergence for both the OS and NS star formation schemes. Among the models considered, OS-RCO-mw shows the best convergence in $\bar\eta_M$, even better than NS-RCO-gs.

\paragraph{Energy loading factor}

The middle of Figure~\ref{fig:eta_all} shows $\bar\eta_E$ and the corresponding relative errors at $r=4\,\rm kpc$. For the reference OS-FNG-mw model, $\bar\eta_E$ increases substantially from $\sim 0.05$ to $\sim 0.35$ as the resolution coarsens from $4$ to $32\,\rm pc$, resulting in poor convergence, with RE4 increasing from $\sim70\%$ at $\Delta x=8\,\rm pc$ to above $100\%$ at $\Delta x\geq16\,\rm pc$. In contrast, $\bar\eta_E$ in NS-RCO-gs fluctuates within $\sim 0.03$--$0.12$ across $4$ to $32\,\rm pc$, showing a weaker resolution dependence than OS-FNG-mw and the best convergence among all models, with RE4 ranging from $\sim70\%$ to $\sim80\%$ across the full resolution range.

Modifying only the sink control volume (HS-FNG-mw), the star formation efficiency (HS2-FNG-mw), or both (NS-FNG-mw) does not improve the convergence. The OS-RCO-mw model, which modifies only the SN coupling radius while retaining the original star formation prescription, improves the convergence relative to OS-FNG-mw, but RE4 remains above $\sim100\%$ at $\Delta x\geq16\,\rm pc$. NS-RCO-mw provides further improvement, with RE4 below $\sim50\%$ at $\Delta x\leq16\,\rm pc$, although it rises above $\sim100\%$ at $\Delta x=32\,\rm pc$.

\paragraph{X-ray emission efficiency}

The right panel of Figure~\ref{fig:eta_all} shows $\bar\epsilon_{L_X}$ and corresponding relative errors. For the reference OS-FNG-mw model, $\bar\epsilon_{L_X}$ increases from $\sim0.01$ to $\sim0.02$ as the resolution coarsens from $4$ to $32\,\rm pc$. This increase results from $\langle\dot{E}_{\rm SN}(t)\rangle_{5-25\,\rm Myr}$ decreasing by $\sim88\%$, compared with a $\sim70\%$ decrease in $\langle L_X\rangle_{10-30\,\rm Myr}$. Consequently, RE4 increases from $\sim30\%$ at $\Delta x=8\,\rm pc$ to $\sim100\%$ at $32\,\rm pc$.

In contrast, $\bar\epsilon_{L_X}$ in the fiducial NS-RCO-gs model fluctuates around $\sim0.008$, with RE4 remaining below $20\%$ across the full resolution range. The HS-FNG-mw and HS2-FNG-mw models show no improvement in convergence than OS-FNG-mw, whereas modifying the SN coupling scheme, either alone or together with the updated star formation prescription, substantially improves convergence in OS-RCO-mw, NS-RCO-gs, and NS-RCO-mw.

On the other hand, we have tested the sensitivity of our results to the averaging time windows and offset intervals adopted in Equations~\ref{eqn:ave_mass_load} , ~\ref{eqn:ave_ene_load} and ~\ref{eqn:ave_x_eff}. Increasing the averaging windows and changing the offset intervals does not change the overall trends in resolution convergence among different models, although the resulting RE4 values change moderately for $\bar\epsilon_{L_X}$, and mildly for $\bar\eta_M$ and $\bar\eta_E$.

In summary, modifying the star formation prescription alone does not improve the resolution convergence of the mass and energy loading factors or X-ray emission efficiency. Changing the SN feedback coupling radius from $3\Delta x$ to $R_{\rm cool}$ generally improves convergence, while combining the updated star formation and SN coupling schemes provides further improvement. Our fiducial NS-RCO-gs model achieves RE4 below $\sim50\%$, $\sim80\%$, and $\sim20\%$ for the averaged mass loading factor, energy loading factor, and X-ray emission efficiency, respectively, across $\Delta x=4$--$32\,\rm pc$.

\section{Discussions}
Resolution convergence remains a challenge for hydrodynamical simulations of galaxy evolution. As sub-grid prescriptions for baryonic physics, such as star formation and stellar feedback, become increasingly complex, ensuring that macroscopic outcomes remain independent of numerical resolution is essential for physical reliability. Consequently, a significant body of high-resolution idealized simulations has emerged to systematically investigate how the state of the interstellar medium (ISM), star formation rates, and galactic outflow properties scale with resolution and its associated sub-grid parameters (e.g., \citealt{2018MNRAS.478..302S, Hu2019, 2020MNRAS.495.1035S, 2024A&A...691A.231D}).

Our preferred models, NS-RCO-mw and NS-RCO-gs, successfully reproduce the observed starburst and multiphase outflow properties of M82 while maintaining robust numerical convergence. As the resolution is degraded from $4\,\mathrm{pc}$ to $32\,\mathrm{pc}$, they keep the relative errors below $25\%$ for star formation and total kinetic energy, below $\sim 40\%$ for the peak cool-phase mass outflow rate, below $\sim 50\%$ for the warm phase, and below $\sim 75\%$ for the hot phase. The best-performing configuration also limits the relative error in the integrated X-ray emission to below $\sim 40\%$.

Beyond these integrated quantities, we examine the convergence of normalized wind properties, the mass loading factor, energy loading factor, and X-ray emission efficiency, which more directly probe wind-driving efficiency by removing the dependence on the overall star formation and SN energy budgets. Combining the updated star formation and SN coupling schemes, our fiducial NS-RCO-gs model substantially improves the convergence of these properties relative to the original OS-FNG-mw model, with RE4 below $\sim50\%$, $\sim80\%$, and $\sim20\%$ for the averaged mass loading factor, energy loading factor, and X-ray emission efficiency, respectively, across $\Delta x=4$--$32\,\rm pc$.

In the following sections, we compare our approach with previous studies and identify the key ingredients required to achieve converged predictions for star formation and supernova-driven winds in galaxy-scale simulations.

\subsection{convergence on star formation}
Methodologies for modeling star formation vary significantly based on spatial and temporal resolution. High-resolution simulations ($\Delta x \lesssim \text{GMC scales}$) often utilize the sink particle method \citep{Federrath2010}, which resolves gravitational collapse and gas accretion onto dense cores with sub-Myr temporal precision. In contrast, simulations that cannot resolve individual GMCs typically employ sub-grid star formation laws. These models convert gas into stars based on a prescribed efficiency once a density threshold is reached, mimicking empirical relations between gas density and star formation at hundred-parsec to kiloparsec scales.

For the sink particle method, convergence is primarily governed by the formation criteria, the definition of the control volume, the accretion rate, and the internal star formation efficiency. While formation criteria, such as Jeans instability, convergent flow, and non-overlapping conditions, are relatively standardized \citep{Federrath2010}, resolution convergence remains highly sensitive to how accretion and SFE are handled. 

Recent studies have attempted to mitigate resolution dependence by scaling these parameters. For instance, \citet{2018MNRAS.478..302S} adopted a resolution-dependent Jeans length, while \citet{2021MNRAS.506.2199G} demonstrated that star formation histories become insensitive to resolution when the gas mass resolution reaches $\Delta m < 0.01\,\rm M_{\odot}$. However, at the intermediate resolutions typical of galaxy-scale simulations, the intrinsic SFE assigned to a sink particle remains a major source of divergence.

High-resolution simulations have shown that the star formation efficiency at GMCs scale is not universal, but instead scales with cloud surface density, $\Sigma_{\rm cl}$, approximately following $\epsilon \propto \Sigma_{\rm cl}^{n}$ \citep[e.g.,][]{Kim_2018, li2019disruption, 2021MNRAS.506.5512F}. Early stellar feedback further regulates the SFE by dispersing molecular clouds before all of their gas can be converted into stars \citep{li2019disruption}. When simulations lack the resolution to capture these internal GMC processes, the SFE must be specified through a sub-grid prescription for sink particles. Because such prescriptions often vary substantially and are frequently calibrated for a particular resolution, achieving a converged SFRs remains a challenge for galaxy simulations.

In cosmological and isolated galaxy simulations, where the typical resolution is coarser than $\sim 10\,\mathrm{pc}$,
the internal structure of the ISM and the detailed physics of star formation remain unresolved. In this regime, star formation is highly sensitive to the adopted density threshold and sub-grid star formation prescription. Cosmological simulations commonly employ thresholds of $n_0 \sim 0.1\,\mathrm{cm^{-3}}$ to identify star-forming gas \citep{2015MNRAS.450.1937C,2018MNRAS.473.4077P}, whereas isolated galaxy simulations at high resolution typically adopt much higher values, ranging from $10^2$ to $10^5\,\mathrm{cm^{-3}}$ \citep{2018MNRAS.480..800H,Emerick2019,2019MNRAS.489.4233M,2021MNRAS.501.5597G,2024A&A...691A.231D}.

Once the threshold is met, a Kennicutt-Schmidt (KS) relation is typically used to convert gas into star particles. However, the resulting star formation intensity is highly sensitive to the chosen threshold and free parameters within the model; a mere factor of two \citep{Emerick2019} or few \citep{2018MNRAS.478..302S, Hu2019} change in resolution can lead to significant divergence in results. While large-scale cosmological projects (e.g., TNG, EAGLE) calibrate these parameters against global observables like the galaxy stellar mass function, applying these calibrated values to different resolutions often results in failure. For isolated galaxy simulations, this challenge is even more acute, as there are few observational constraints beyond the KS law to guide parameter recalibration as resolution varies.

Building upon the sink particle framework, our ``NS'' model achieves robust convergence across the 4--32 pc resolution range in M82-like systems with high gas surface density and star formation rates, a critical gap between high-resolution isolated galaxy models and large-scale cosmological simulations. Notably, the resulting starburst properties align well with observations of M82. This improved convergence compared to \citet{li2025revisiting} is driven by several key advancements:

First, we adopt well-justified formation criteria for sink particles, including Jeans instability, convergent flow, gravitational potential minimum and non-overlapping conditions \citep{Federrath2010}. Second, we utilize a fixed physical control volume ($\sim20$ pc, close to the typical size of GMCs) for gas accretion, rather than a volume scaled to the grid size. This ensures that the gas reservoir available to each sink particle remains physically consistent and largely independent of numerical resolution.

Finally, our star formation scaling relations employ a resolution- and density-dependent star formation efficiency per local free-fall time. By calibrating the characteristic surface density parameter, $\Sigma_{0}$, as the grid coarsens that broadly consistent with observed densities of interstellar gas structures at comparable physical scales, we further mitigate the divergence of SFRs typically caused by assuming a fixed efficiency parameter across varying resolutions/density.

\subsection{Convergence on SN Feedback Coupling and Galactic Outflows}
\label{sec:disSN}

Stellar feedback, primarily in the form of supernova (SN) explosions, is a fundamental driver of galaxy evolution. Its impact spans a vast dynamic range: from the sub-parsec evolution of individual supernovae remnants (SNRs) to the maintenance of the turbulent interstellar medium (ISM) and the launching of kiloparsec-scale galactic outflows.
In numerical simulations, various sub-grid models are employed to couple SN energy with the neighboring ISM. Depending on the specific framework, the energy released is injected into the environment as pure thermal energy, pure kinetic energy, or a combination of both.

In its simplest form, SN feedback energy is modeled by injecting $10^{51} \rm\, erg$ of thermal energy. While effective in high-resolution simulations that resolve the Sedov-Taylor phase (\citealt{Kim_2015, Hu2019, Emerick2019, 2020MNRAS.495.1035S, 2021MNRAS.501.5597G, 2024A&A...691A.231D}), this approach often fail at coarser resolutions. If the cooling radius is unresolved, the thermal energy is radiated away before it can be converted into kinetic energy. This ``over-cooling'' problem causes significant divergence in star formation and wind properties as resolution changes, leading to marked discrepancies between different SN feedback schemes  (\citealt{2018MNRAS.478..302S, Hu2019, 2022ApJS..262....9O}).

Severe radiative losses occur when SN energy is injected into excessive gas mass, heating it only to the cooling curve peak ($\sim 10^5$ K) where cooling times are shortest \citep{2012MNRAS.426..140D}. To compensate this issue, some models artificially delay cooling (\citealt{2006MNRAS.373.1074S, 2013MNRAS.429.3068T, 2015MNRAS.454...83W}) or employ stochastic thermal feedback that heats only a fraction of the gas to higher temperatures, $\sim 10^{7.5} \rm \,K$ \citep{2012MNRAS.426..140D}.
Others inject multiple SNe simultaneously to reach higher temperatures at the cost of temporal frequency \citep{2017MNRAS.470L..39F}. However, these methods often show relatively poor convergence in the SFR, ISM structure, and outflow properties across resolutions, and may distort the multiphase structure of the ISM and galactic winds \citep{2018MNRAS.478..302S, 2022MNRAS.514..249C}.

To mitigate over-cooling, kinetic injection schemes deliver explosion energy as direct momentum kicks to the surrounding gas \citep{2008A&A...477...79D, 2008MNRAS.387.1431D}. However, both thermal and kinetic components are essential for a realistic ISM, as they respectively drive the hot and cool phases of galactic outflows \citep{Hu2019, 2022ApJS..262....9O}. Consequently, many simulations employ hybrid models that split SN energy into thermal and kinetic fractions (\citealt{2015ApJ...809...69S, 2018MNRAS.473.4077P, 2019MNRAS.486.2827D, 2023MNRAS.523.3709C}). While these models can replicate global galaxy statistics, their specific energy-partitioning ratios vary significantly across the literature. These discrepancies lead to inconsistent ISM and outflow properties \citep{2015ApJ...809...69S,2020MNRAS.494.3971M}, and ultimately hinder resolution convergence. 

A critical test for any supernova feedback model is the convergence of the predicted turbulent ISM, star formation activity, and large-scale outflows across resolutions. When the grid scale is fine enough to resolve the early stages of SNR, thermal, kinetic, and hybrid models generally yield converged global outflow properties, such as mass-loading factors and average velocities. However, once the resolution fails to capture these early phases, significant discrepancies emerge in both star formation and wind properties across different feedback implementations \citep{2018MNRAS.478..302S, Hu2019}.

To address this, an upgraded class of mechanical feedback models has been proposed based on high-resolution simulations of SNR evolution. These studies demonstrate that the initial momentum from the free-expansion phase is boosted by a factor of $7$--$10$ during the pressure-driven snowplow phase (\citealt{2014MNRAS.445..581H,2014ApJ...788..121K, Kim_2015, 2015MNRAS.450..504M}). In this class of feedback model, the terminal momentum is assigned to the surrounding ISM based on the number of SN events and local gas density, while the remaining energy is injected in thermal mode. By ensuring the feedback captures the physical solution of the Sedov-Taylor phase regardless of resolution, this method has been widely adopted to improve the resolution convergence of star formation and outflow properties in galaxy-scale simulations (e.g., \citealt{2018MNRAS.478..302S, Hu2019}).

At intermediate resolutions ($\sim 10$--$100\rm\,pc$), typical for next-generation cosmological simulations, such mechanical feedback models are very helpful for ensuring resolution convergence \citep{2018MNRAS.480..800H,2022ApJS..262....9O}. Our approach builds upon the resolution-dependent ``compound'' feedback model from \citet{kim2017three}, which determines the coupling scheme via the ratio of the feedback deposit mass to the shell formation mass, $R_{M}$. This ratio is sensitive to both the size of the injection zone and the local ISM density. 

When $R_{M} > 1$, the Sedov-Taylor phase is unresolved or the SNR has already entered the radiative phase; in this regime, we employ pure momentum injection. Conversely, $R_{M} < 1$ indicates that the Sedov-Taylor stage is well-resolved. In a slight departure from the original $72\%/28\%$ thermal-to-kinetic split in \citet{kim2017three}, we adopt an equal ($50\%/50\%$) energy partition for the $R_{M} < 1$ case to better balance the driving of hot and cool gas phases (see Appendix \ref{sec:sn_feedback} for details).

\begin{figure*}[!htbp]
    \centering
    \includegraphics[width=0.75\textwidth]{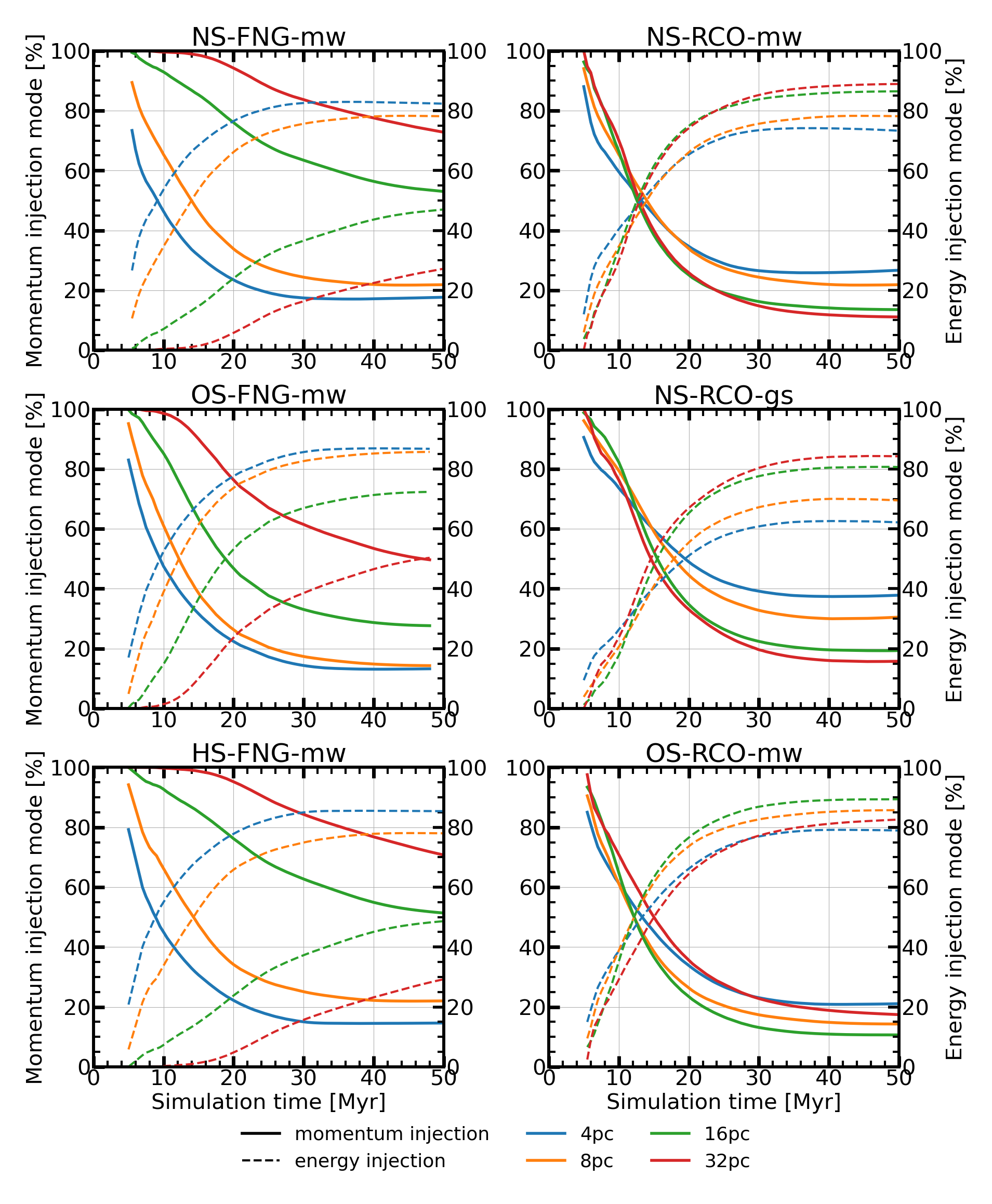}
    \caption{Proportion of Momentum and energy injection of SNe feedback vs. simulation time. For a simulation, the sum of the proportions of mode 1 and mode 2 is 100\%. The titles represent different simulation settings for each group, and the lines inside at four different resolutions.}
    \label{fig:mode1}
\end{figure*}

Furthermore, our model injects SN feedback into a volume defined by a physical scale close to the cooling radius, $R_{cool}$, the scale at which an SNR transitions from the energy-conserving to the momentum-conserving phase, rather than a volume tied to a fixed number of grid cells. We now explore how the selection of the feedback deposit zone influences coupling within our resolution-dependent framework. Figure \ref{fig:mode1} illustrates the resulting fraction of feedback modes over time across varying resolutions; specifically, it shows the fraction of events utilizing pure momentum injection ($R_{M} > 1$, solid lines) versus hybrid energy injection (dashed lines).

When the first supernovae (SNe) appear at $t \approx 5\rm\,Myr$, feedback is handled predominantly via momentum injection across all groups. Initially, star formation is concentrated in high-density central gas clumps where the cumulative feedback is too weak to significantly alter the dense surrounding ISM. However, as feedback energy accumulates between $t=5$ and $t=20\rm\,Myr$ (see Figure \ref{fig:sfr_Moutcool_35}), the starburst region thermalizes, driving the formation of hot superbubbles \citep{Wang2026}. This transition lowers the local ISM density, causing the fraction of $R_M > 1$ (momentum-mode) events to decrease.

The decline of the momentum mode is remarkably consistent across resolutions in the ``RCO'' scheme. In contrast, the ``FNG'' scheme shows a strong resolution dependence: because its injection radius is set to $3\Delta x$, coarser resolutions ($\Delta x \geq 16$ pc) create excessively large coupling volumes. This leads to over-sampling of gas mass, causing SNRs in the coarser ``FNG'' runs to remain in the momentum-injection regime even after $t\sim20\,\mathrm{Myr}$, with inter-resolution differences reaching $\sim 35\%$–$60\%$ by $t=50\,\mathrm{Myr}$ (left column). In comparison, ``RCO'' approach yields a more resolution-independent transition between feedback modes, reducing discrepancies across resolutions to $\sim 10\% -25\%$ at $t=50\rm\,Myr$ (right column).

Note that, for the $4\,\mathrm{pc}$ runs, the differences on the injection mode between the ``RCO'' configuration and ``FNG'' configuration are much narrower than coarser resolutions. The main reason is that, at this resolution, SN feedback is well resolved in most case and the details of the injection scheme become unimportant (\citet{2017ApJ...834...25K}). However, the ``RCO'' configuration adopts a larger feedback injection radius ($r_{\rm snr}=4\Delta x$) than the fixed-grid ``FNG'' configuration ($r_{\rm snr}=3\Delta x$). As a result, the fraction of SNe treated in the momentum-injection mode is modestly higher in the $4\,\mathrm{pc}$ ``RCO'' runs than in the corresponding ``FNG'' runs all the time, as shown by the blue curves in Figure~\ref{fig:mode1}.

Moreover, for the ``RCO'' configuration, the $4\,\mathrm{pc}$ runs in all simulation groups exhibit a slower decline in the momentum-injection fraction than their coarser-resolution counterparts after $t\sim6\,\mathrm{Myr}$, and eventually become the highest by $t\sim10-20\,\mathrm{Myr}$ despite starting from the lowest level at $t\sim5\,\mathrm{Myr}$. At first glance, this trend appears counterintuitive, since higher resolution should better resolve the Sedov--Taylor phase and thus favor energy injection. However, the effect is likely driven by the stronger central star formation in the $4\,\mathrm{pc}$ runs (Figure~\ref{fig:mw3sfdis}), which maintains a denser ambient environment around many SN events. As a result, SNe occurring after $t\sim10\,\mathrm{Myr}$ in 4 pc run are more likely to satisfy $R_M>1$ than at coarser resolution and therefore enter the momentum-injection regime, offsetting the expected increase in the energy-injection fraction due to improved numerical resolution.

Therefore, while Figure \ref{fig:mode1} demonstrates improved convergence of the ``RCO'' configuration over the 4--32 pc range, at resolutions finer than 4 pc the optimal approach would be directly resolve the early phases of SNR evolution. On the other hand, the improved resolution convergence of feedback modes in the ``RCO'' groups can explains their more consistent integrated and normalized outflow properties (Section \ref{sec:results}). While our model includes clustered SN feedback to enhance momentum contribution \citep{Gentry17}, the ``FNG'' scheme’s reliance on grid-dependent volumes leads to an artificially too high proportion of momentum injection at coarser resolutions. This over-injection is evident in the NS-FNG-mw and HS-FNG-mw groups, which exhibit inflated mass outflow rates and X-ray luminosities in coarser runs (see Figures \ref{fig:Mout_diff}, and \ref{fig:xray_error}).

To accurately reproduce M82-like starburst properties while ensuring resolution convergence, we find that a combination of key strategies is essential: the resolution-dependent compound feedback framework \citep{kim2017three}, clustered SN momentum enhancements, an injection zone scaled to $R_{\rm cool}$, and a Gaussian-weighted assignment scheme.

\subsection{limitation of our study}
While our model demonstrates improved convergence, several limitations remain. First, the sink particle and feedback coupling methods are still constrained by the grid scale. Our approach uses a control volume and feedback deposit zone ($\sim 20$ pc) comparable to the size of GMCs and $R_{\text{cool}}$. At coarser resolution, it reduces to a single-cell implementation, reintroducing resolution dependence. Besides, for NS-RCO-gs configuration, the adopted Gaussian kernel widths, $\sigma=4,3,2$ and $1\Delta x$ for $\Delta x = 4,8,16,32\,\rm pc$, respectively, do not scale uniformly with resolution. These values were chosen to simply match the adopted injection radius. As a result, the Gaussian profile becomes only marginally resolved at $\Delta x=32 \rm\, pc$, and the deposition approaches a nearly uniform weighting.

Second, although key properties converge within the $4$–$32\,\mathrm{pc}$ range, non-negligible discrepancies persist in several diagnostics. In particular, while the total and cool-phase outflow rates converge to within $\sim 40\%$, the warm and hot phases remain less well converged, with relative errors reaching $\sim 50\%$ and $\sim 75\%$, respectively. Also, the relative errors of time averaged mass loading factor and energy loading factor is about $50\%$ and $80\%$. Additional differences are also present in the timing of the SFR peak and in the central stellar spatial distribution (see Figure~\ref{fig:mw3sfdis}).

Furthermore, the simulation duration is relatively short, as our primary goal was to test convergence during the peak starburst phase of M82-like galaxies. Longer timescales and larger volumes are necessary to track the long-term evolution of outflows and their impact on galaxy (\citealt{2017MNRAS.470L..39F, 2024MNRAS.52710095V}). Also, our resolution does not fully capture small-scale mixing in the outer outflow regions, which may influence late-stage convergence (\citealt{2017MNRAS.470L..39F, schneider2020physical, 2024MNRAS.52710095V}). Meanwhile, a more comprehensive model of ``early'' stellar feedback (e.g., photoionization and stellar winds) would likely enhance the robustness of our results.

Finally, while our framework is converged for tens-of-parsec scales typical of low-mass, M82-like starburst galaxies, extending robust convergence across all galaxy types and scales remains an open challenge.

\section{Summary}
\label{sec:summary}

To mitigate resolution dependence in simulations of starburst and outflow in M82-like galaxy, we have introduced and evaluated two critical upgrades to the star formation and supernova (SN) feedback modules within the framework developed in \cite{Wang2026} and \cite{li2025revisiting}, which is implemented in the Athena++ code. Our refined model reproduces the observed starburst and multiphase outflow properties of M82 and achieves good convergence across $4$–$32\,\mathrm{pc}$ resolution, with relative errors below $\sim 20\%$ for stellar mass and total kinetic energy, below $\sim 40\%$ for the cool-phase mass outflow rate, and below $\sim 25\%-40\%$ for the integrated X-ray emission. The warm and hot outflow phases remain less well converged, with relative errors of $\sim 50\%$ and $\sim 75\%$, respectively. Our refined model also improve the convergence on normalized wind properties, with relative errors below $\sim 50\%$, $\sim 80\%$ and $\sim 20\%$ for the time averaged mass loading factor, energy loading factor and X-ray emission efficiency. Our key findings are summarized below: 

\begin{enumerate}
\item By anchoring the sink particle accretion volume to a physical scale of $\sim 20\rm\,pc$, approximately the typical size of Giant Molecular Cloud (GMC), rather than a fixed number of grid cells, we decouple star formation from the numerical grid. Combined with a resolution-dependent surface density threshold that generally aligns with observational trends, this scheme ensures the total stellar mass formed varies by less than $\sim 20\%$ across the tested resolution range, against to $\sim 60\%$ with the previous model. The updated prescription also produces a cumulative stellar mass in better agreement with observations at the highest resolution of $4\,\rm pc$.

\item Building on the compound feedback model of \cite{kim2017three}, we demonstrate that setting the SN injection radius close to the SNR cooling radius ($R_{\rm cool}$) can effectively stabilizes the compound SN feedback prescription, i.e., reduce the discrepancies on momentum/thermal injection mode and outflow properties across resolutions. Across resolutions of 4–32 pc, the total kinetic energy varies by less than $\sim20\%$, while the cool-phase mass outflow rate and integrated hot-phase X-ray luminosity vary by $\sim25$--$40\%$. In particular, the variation in the cool-phase mass outflow rate is reduced from more than $100\%$ in the previous model to below $\sim40\%$. However, the warm and hot mass outflow rates remain less well converged, with relative errors of approximately $\sim50\%$ and $\sim75\%$, respectively. These results underscore that breaking the artificial scaling between feedback zones and numerical resolution is essential for achieving convergence in feedback coupling and galactic winds.

\item To separate the resolution dependence of the feedback driving budget from the intrinsic wind-launching efficiency, we further analyze the normalized wind properties, including mass loading factors, energy loading factors, and X-ray emission efficiency. The fiducial NS-RCO-gs model shows improved convergence in these quantities compared with the original scheme. Across resolutions of 4–32 pc, the relative errors of time averaged mass loading factors, energy loading factor and X-ray emission efficiency remain below $\sim50\%$, $\sim80\%$ and $\sim 20\%$, respectively. These results demonstrate that the revised prescriptions improve the numerical convergence of both the integrated wind properties and the efficiency of stellar feedback in driving multiphase galactic outflows.

\end{enumerate}

These optimizations provide an effective framework for modeling star formation and supernova feedback in M82-like starburst galaxies over resolutions of $4$--$32\,\mathrm{pc}$. By anchoring the prescriptions to their characteristic physical scales and accounting for resolution-dependent ISM density trends, the model substantially reduces the sensitivity of key galaxy and outflow properties to numerical resolution. While calibrated and tested specifically for M82-like systems, this framework serves as a useful testbed for developing more robust sub-grid models for galaxy formation simulations. Extending and validating the approach across a broader range of galaxy types remains an important goal for future work. 

\begin{acknowledgments}
We are grateful to the anonymous reviewer for their valuable comments and suggestions, which have helped improve the manuscript. This work is supported by the National Natural Science Foundation of China (NFSC) through grant 12595314, and National SKA Program of China (grant Nos. 2025SKA0150100 and 2025SKA0150104). WSZ is supported by NSFC grants 12173102 and 11733010. The simulations carried out in this work was completed on the HPC facility of the School of Physics and Astronomy, Sun, Yat-Sen University.

\end{acknowledgments}

\appendix

\section{star formation efficiency formula at sink/clump scale}

On galactic scales, observations \citep[e.g.,][]{1998ARA&A..36..189K} reveal a robust correlation between the star formation surface density, $\Sigma_{\rm SFR}$, and the gas surface density, $\Sigma_{\rm gas}$, typically expressed as:

\begin{equation}
\label{eq:SFRtoGAS}
\Sigma_{\rm SFR} \propto \Sigma_{\rm gas}^{N},
\end{equation}
where $N$ is the power-law index. Consequently, the characteristic star formation timescale at galactic scales can be defined as:

\begin{equation}
\label{eq:tsf_NEW}
t_{\rm sf, gal} = \frac{\Sigma_{\rm gas}}{\Sigma_{\rm SFR}} \propto \Sigma_{\rm gas}^{1-N}.
\end{equation}

Observations demonstrate that this scaling relation between $\Sigma_{\rm SFR}$ and $\Sigma_{\rm gas}$ persists down to sub-kpc scales \citep[e.g.,][]{2007ApJ...671..333K, 2008AJ....136.2846B, 2013AJ....146...19L}. 

\citet{2005ApJ...630..250K} proposed a turbulence-regulated model to explain the observed $\Sigma_{\rm SFR}$–$\Sigma_{\rm gas}$ scaling and further proposed a universal star formation law spanning scales from individual molecular clouds to entire galaxies, characterized by a near-constant SFE of a few percent per local free-fall time. 

More recent observations have investigated the scaling relations between the $\Sigma_{\rm SFR}$ and $\Sigma_{\rm gas}$ within Milky Way molecular clouds, spanning scales from entire GMCs down to their dense clumps and cores ($\sim 0.1$--$10$ pc) \citep[e.g.,][]{2009ApJS..181..321E, 2010ApJ...723.1019H, 2014ApJ...782..114E, 2021ApJ...912L..19P, 2025MNRAS.540.2377R}. Several of these studies report power-law relations of the form $\Sigma_{\rm SFR} \propto \Sigma_{\rm gas}^{N}$ with $1 \lesssim N \lesssim 2$, or $\Sigma_{\rm SFR} \propto \epsilon_{\rm ff} (\Sigma_{\rm gas}/t_{\rm ff})^{N}$ with $N \simeq 0.8$--$1.0$ and a star formation efficiency of $\epsilon_{\rm ff} \approx 0.026$--$0.16$ per free-fall time \citep{2010ApJ...723.1019H, 2021ApJ...912L..19P, 2025MNRAS.540.2377R}.

In this work, we implement a star formation prescription that bridges the scales of individual GMC clumps ($\sim 1$--$20$ pc) and galactic scale, informed by the theoretical and observational findings discussed above. We determine the functional form of the $\epsilon_{\rm ff}$ by requiring that the simulated star formation properties statistically recover the empirical star formation law observed from galactic down to sub-GMC scales, $\Sigma_{\rm SFR,clump} \propto \Sigma_{\rm clump}^{N}$. 

For a local star formation clump (both include sink particle and its host cell), the surface density is estimated as $\Sigma_{\rm clump} = M_{\rm clump} / \Delta x^2$ with an approximate of characteristic radius close to the grid resolution, $R_{\rm clump} \sim \Delta x$. Since most sink particle in our simulations formed in cold gas near a temperature region from $300\rm\,K$ to a few $10^{3}\rm\,K$, we model the star formation surface density at the cold clump (grid) scale as:

\begin{equation}
\label{eq:sfr_assume}
\Sigma_{\rm SFR,clump} = \frac{\Sigma_{\rm clump}}{t_{\rm sf,clump} }=\epsilon_{\rm ff,clump} \frac{\Sigma_{\rm clump}}{ t_{\rm ff,clump}},
\end{equation}

Given that the grid resolution $\Delta x$ remains constant throughout a specific simulation, the volume density and surface density are related by:

\begin{equation}
\label{eq:tfftosigma}
\rho_{\rm clump} \propto \frac{\Sigma_{\rm clump}}{R_{\rm clump}} \propto \frac{\Sigma_{\rm clump}}{\Delta x} \propto \Sigma_{\rm clump}.
\end{equation}

It follows from the definition of the free-fall time that:
\begin{equation}
t_{\rm ff,clump} \propto \rho_{\rm clump}^{-0.5} \propto \Sigma_{\rm clump}^{-0.5}.
\end{equation}

Substituting these relations into Equation \ref{eq:sfr_assume}, we find that if $\epsilon_{\rm ff}$ scales as:
\begin{equation}
\label{eq:eff_rela}
\epsilon_{\rm ff,clump} \propto \frac{1}{\Sigma_{\rm clump}^{1.5-N}},
\end{equation}
the empirical star formation law is recovered at the clump scale.

Thus, to ensure numerical convergence across varying spatial scales, we adopt the following density-dependent formulation for the efficiency per local free-fall time $\epsilon_{\rm ff}$:

\begin{equation}
\epsilon_{\rm ff,clump} = \epsilon_{\rm 0,ff} \left( \frac{\Sigma_{0}}{\Sigma_{\rm clump}} \right)^{1.5-N},
\end{equation}
where $\epsilon_{0,\rm ff}$ and $\Sigma_{0}$ are characteristic values given in Section 2.3. 

\section{Supernovae feedback}
\label{sec:sn_feedback}
In this Appendix, we provide a detailed description of the supernova (SN) feedback module. Our basic framework follows the compound feedback model of \cite{kim2017three}. In \cite{li2025revisiting}, a spherical feedback region with a radius fixed to three grid cells, $r_{\rm snr}=3\Delta x$, was utilized. In this work, we implement an improved scheme where the radius of the SN injection zone is anchored to the physical cooling radius ($R_{\rm cool}$), which is approximately $20$ pc for a mean ISM density of $1\rm\,cm^{-3}$. To maintain this physical scale across different grid sizes, we set $r_{\rm snr}$ to $4\Delta x$, $3\Delta x$, $2\Delta x$, and $\Delta x$ for resolutions of $\Delta x=4$, 8, 16, and 32,pc, respectively.

Based on the gas properties within the injection zone, we apply three distinct prescriptions to handle supernova feedback. At each time step, we calculate the total mass, $\sum M_{i,j,k}$, and volume, $\sum V_{i,j,k}$, of the grid cells within the deposit zone. For any star particle triggering a supernova event, we determine the ejected mass, $M_{\rm ej}$, energy, $\Delta E_{\rm sn}$, and metal mass, $M_{\rm metal,ej}$, following the methodology in Section 2.2 of \cite{li2025revisiting}. The mean density of the SNR region is then computed as $\rho_{\rm snr} = (\sum M_{i,j,k} + M_{\rm ej}) / \sum V_{i,j,k}$. Following \cite{Kim_2015}, the mass at which shell formation occurs is defined as:
\begin{equation}
M_{\rm sh} = 1679\,M_{\odot} \left( \frac{n_{\rm H}}{1\,\rm cm^{-3}} \right)^{-0.26},
\end{equation}
where the hydrogen number density is $n_{\rm H} = \rho_{\rm snr}/(\mu_{\rm H}m_{\rm H})$ with $\mu_{\rm H} = 1.427$. Finally, we evaluate the ratio of the total SNR mass to the shell formation mass, $R_{M} = M_{\rm snr}/M_{\rm sh}$, to determine the appropriate feedback coupling mechanism.

When $R_{M} > 1$, the Sedov-Taylor phase of the supernova remnant cannot be fully resolved. In this regime, we inject momentum into the grid cells within the feedback zone. The total terminal radial momentum is determined by:
\begin{equation}
p_{\rm snr} = 2.8 \times 10^{5} \, \text{M}_{\odot} \, \text{km s}^{-1} \left( \frac{n_{\text H}}{1\,\text{cm}^{-3}} \right)^{-0.17} \cdot f_{\rm boost},
\end{equation}
where $f_{\rm boost}$ is a factor accounting for momentum enhancement from clustered supernovae (\citealt{Gentry17}). In all simulations presented here, we adopt $f_{\rm boost}=10$ following \citet{li2025revisiting}. Unlike \citet{li2025revisiting}, who utilized a mass-weighted scheme, this work implements a Gaussian-distribution injection scheme to assign momentum (or energy) and mass to individual grid cells for both $R_{M} > 1$ and $R_{M} < 1$ cases. The specific assignment procedures for these two schemes are detailed below.

\begin{enumerate}
\item Mass weighted: the momentum distribution to individual grid cell is $f_{ijk} = \rho_{ijk} \times V_{ijk} / \sum M_{ijk}$, where ijk represents the cell indices. The momentum increment for each grid cell in the x, y, z directions is then calculated as $\Delta \overset{\to }p_{ijk} = f_{ijk} \times(\Delta M_{ej}\times \overset{\to }v_{p}+ p_{snr} \times \overset{\to }{n})$, where $\overset{\to }{n}$ is the unit vector pointing from the star particle to the grid cell center, the $\overset{\to }v_{p}$ is the velocity of star particle. The corresponding maximum expected energy increment is $\Delta e_{ijk} = f_{ijk} \times \Delta \rm{E_{sn}}$. 
After updating the momentum for each grid cell, we calculate the change in kinetic energy $\Delta E_{k,ijk}$. If $\Delta E_{k,ijk}$ exceeds $\Delta e_{ijk}$, a correction factor $\alpha=\sqrt{(E_{k0,ijk} + \Delta e_{ijk}) / (E_{k0,ijk} + \Delta E_{k,ijk})}$ is applied, where $E_{k0,ijk}$ is the initial kinetic energy of the grid cell. The final momentum of the grid cell is then updated as $\overset{\to }p_{ijk} = \alpha \times (\overset{\to }p_{0,ijk} + \Delta \overset{\to }p_{ijk})$, where $\overset{\to }p_{0,ijk}$ is the  initial momentum. In cases where $\Delta E_{k,ijk} < \Delta e_{ijk}$, the remaining energy is added to the thermal energy, $\Delta E_{ine,ijk} = \Delta e_{ijk} - \Delta E_{k,ijk}$.

\item In the Gaussian-weighted injection scheme, the momentum increment for a grid cell (i,j,k) is defined as: 
\begin{equation} \Delta \vec{p}_{ijk} = \beta \left(M_{\rm ej}\times \vec{v}_{p} + p_{\rm snr} \times \vec{n} \right) \times \exp\left(-\frac{d^{2}}{2\sigma^{2}}\right), \end{equation} where d is the distance from the star particle to the cell center. The normalization factor $\beta$ ensures conservation across the injection region: \begin{equation} \beta = \left[ \sum_{i,j,k} \exp\left(-\frac{d^{2}}{2\sigma^{2}}\right) \right]^{-1}. \end{equation} We adopt a kernel width $\sigma$ of 4,3,2, and $1\times \Delta x$ for resolutions of 4,8,16, and 32 pc, respectively. These values were adopted to match the injection radius at each resolution. The corresponding maximum energy increment is: \begin{equation} \Delta e_{ijk} = \beta \Delta E_{\rm sn} \exp\left(-\frac{d^{2}}{2\sigma^{2}}\right). \end{equation} To ensure strict conservation, we apply a correction factor $\alpha$ to the momentum or redistribute the residual energy to the thermal component if the realized kinetic energy $\Delta E_{k,ijk}$ is less than the assigned energy budget $\Delta e_{ijk}$.
\end{enumerate}

When $R_{M} < 1$, following the original scheme in \cite{kim2017three}, we partition the total SN energy, $\Delta \rm{E_{sn}}$, equally into kinetic and thermal components. The blast wave velocity $v_{sn}$ at the moment of explosion is defined as: 
\begin{equation} 
v_{\rm sn} = \sqrt{2.0\times(\frac{\Delta E_{\rm sn}}{M_{\rm ej}})/2}. \end{equation}

For each grid cell within the injection zone, the momentum increment along the $x, y, \text{ and } z$ directions is then calculated as follows:

\begin{equation}
    \Delta p_{x,ijk}=\Delta m_{ijk}\times(v_{px} + \frac{x_{ijk}-x_{p}}{d}v_{sn}),
\end{equation}

\begin{equation}
    \Delta p_{y,ijk}=\Delta m_{ijk}\times(v_{py} + \frac{y_{ijk}-y_{p}}{d}v_{sn}),
\end{equation}

\begin{equation}
    \Delta p_{z,ijk}=\Delta m_{ijk}\times(v_{pz} + \frac{z_{ijk}-z_{p}}{d}v_{sn}),
\end{equation}
where $\Delta m_{ijk}$ is the mass increment of gird, $v_{px}, v_{py},v_{pz}, x_{p}, y_{p}, z_{p}$ are the velocity and coordinate of star particle, $d$ is the distance from grid center to star particle. 

The increment of thermal energy is
\begin{enumerate}

\item Mass weighted:

\begin{equation}
    \Delta E_{ine,ijk}=f_{ijk} \times 0.5\Delta E_{sn} 
\end{equation}

\item Gaussian form:

\begin{equation}
    \Delta E_{ine,ijk}= \beta \times 0.5\Delta E_{sn}\times \exp\left(-\frac{d^{2}}{2\sigma^{2}}\right)
\end{equation}

\end{enumerate}

In both the $R_{M} > 1$ and $R_{M} < 1$ regimes, the mass increment is calculated as:

\begin{enumerate}

\item Mass weighted:

\begin{equation}
    \Delta m_{ijk}=f_{ijk} \times M_{ej}
\end{equation}

\item Gaussian form:

\begin{equation}
    \Delta m_{ijk}=\beta \times M_{ej} \times\exp\left(-\frac{d^{2}}{2\sigma^{2}}\right)
\end{equation}

\end{enumerate}

In addition, the metal increment of each grid cell within the SNR also follows the two aforementioned distribution schemes.

\section{Mass outflow rate of cool phase}
\label{sec:Mout_othertime}
\begin{figure*}[!htbp]
    \centering
    \includegraphics[width=0.85\textwidth]{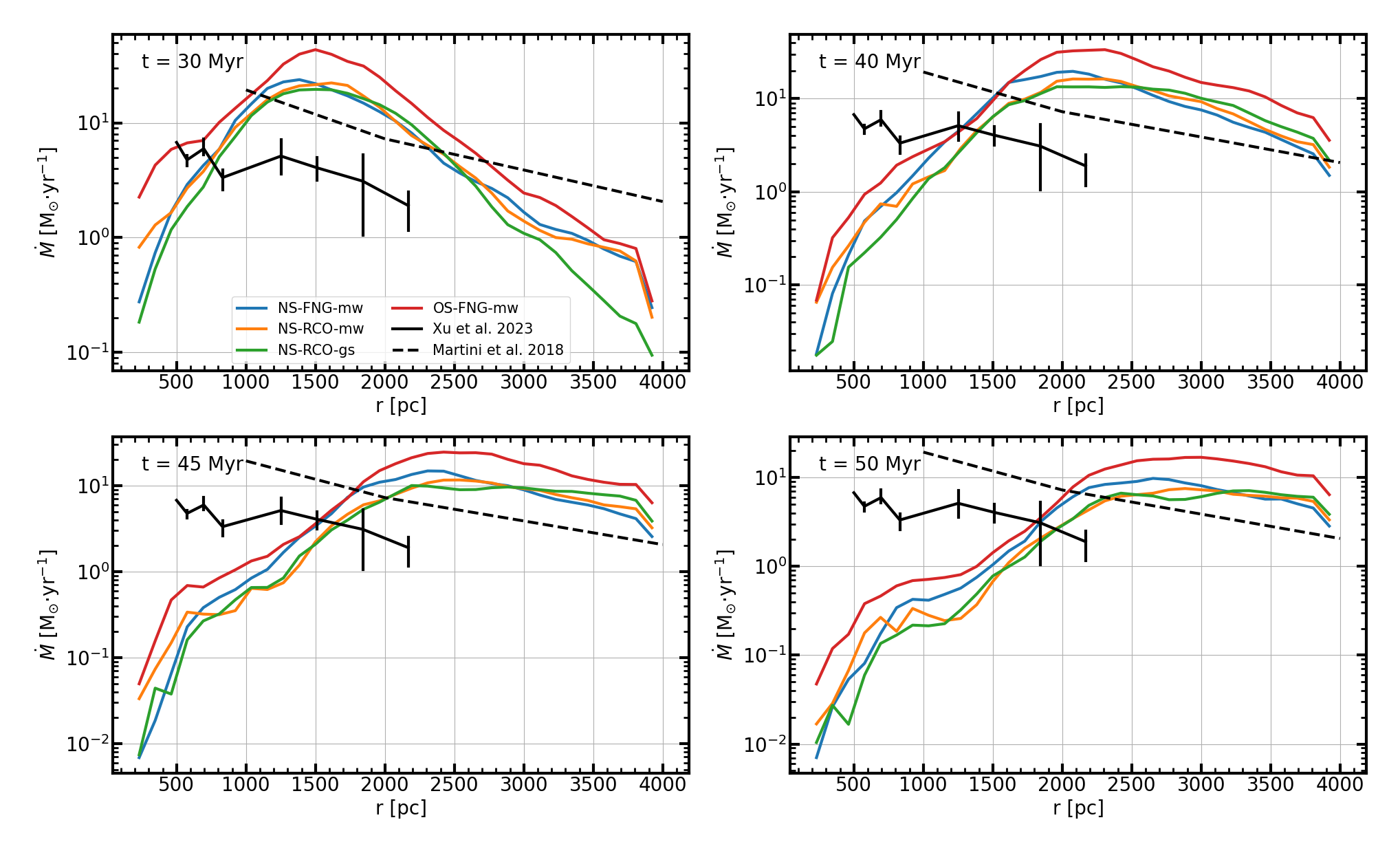}
    \caption{Same as the bottom of Figure \ref{fig:sfr_Moutcool_35}, but measured at $t=30,40,45,50\rm \,Myr$.}
    \label{fig:Mout_multi}
\end{figure*}

Figure \ref{fig:Mout_multi} shows the mass outflow rate of the cool phase measured at $t=30,40,45,50$ Myr in simulations with a resolution of 4 pc.

\begin{contribution}
W.S.Z. conceived the initial research concept and led the writing and submission of the manuscript. X.F.L. contributed to the research design, performed the numerical simulations and data analysis, and prepared the initial draft. T.R.W. and C.H. contributed to the simulation framework, while L.L.F. contributed to the research development and manuscript preparation.


\end{contribution}

%



\software{Athena++ 
         }

\vspace{15mm}



\bibliography{main}
\bibliographystyle{aasjournalv7}



\end{document}